\documentclass[twocolumn]{aastex631}

\newcommand{\flux}{\,erg\,cm$^{-2}$\,s$^{-1}$}
\newcommand{\lum}{\,erg\,s$^{-1}$}

\newcommand{\cm}{\,cm$^{-2}$}
\newcommand{\nh}{$N_\mathrm{H}$}

\shorttitle{Chandra Observation of NGC 1559}
\shortauthors{Ma et al.}

\graphicspath{{./}{figures/}}
\usepackage{graphicx}
\usepackage{multirow}
\usepackage{subfigure}

\begin{document}

\title{\textit{Chandra} Observation of NGC 1559:\\Eight Ultraluminous X-ray Sources Including a Compact Binary Candidate}

\author{Chen-Hsun Ma}
\affiliation{Department of Physics, National Cheng Kung University, Tainan 70101, Taiwan}
\affiliation{Institute of Astronomy and Astrophysics, Academia Sinica (ASIAA), Taipei 10617, Taiwan}
\affiliation{Department of Physics, National Taiwan University, Taipei 10617, Taiwan}

\author{Kwan-Lok Li}
\affiliation{Department of Physics, National Cheng Kung University, Tainan 70101, Taiwan}

\author{You-Hua Chu}
\affiliation{Institute of Astronomy and Astrophysics, Academia Sinica (ASIAA), Taipei 10617, Taiwan}

\author{Albert K. H. Kong}
\affiliation{Institute of Astronomy, National Tsing Hua University, Hsinchu 30013, Taiwan}

\correspondingauthor{Kwan-Lok Li}
\email{lilirayhk@phys.ncku.edu.tw}

\correspondingauthor{Chen-Hsun Ma}
\email{chma@phys.ntu.edu.tw}

\begin{abstract}

Despite the 30-year history of ultra-luminous X-ray sources (ULXs) studies, issues like the majority of their physical natures (i.e., neutron stars, stellar-mass black holes, or intermediate black holes) as well as the accretion mechanisms are still under debate. Expanding the ULX sample size in the literature is clearly a way to help. To this end, we investigated the X-ray source population, ULXs in particular, in the barred spiral galaxy NGC 1559 using a \textit{Chandra} observation made in 2016. In this 45-ks exposure, 33 X-ray point sources were detected within the 2\farcm7 isophotal radius of the galaxy. Among them, 8 ULXs were identified with the criterion of the X-ray luminosity $L_x>10^{39}$\lum (0.3--7~keV). Both X-ray light curves and spectra of all the sources were examined. Except for some low-count spectra that only provide ambiguous spectral fitting results, all the X-ray sources were basically spectrally hard and therefore likely have non-thermal origins.  While no strong X-ray variability was present in most of the sources owing to the relatively short exposure of the observation, we found an intriguing ULX, named X-24, exhibiting a periodicity of $\sim$7500\,s with a detection significance of 2.7$\sigma$. We speculate that it is the orbital period of the system. Roche-lobe over flow and Roche limit are consistent with the speculation. Thus, we suggest that X-24 may be the one of the rare compact binary ULXs, and hence, a good candidate as a stellar-mass black hole.

\end{abstract}


\section{Introduction} \label{sec:intro}

Ultraluminous X-ray sources (ULXs) are off-nuclear point-like sources that radiate X-ray emission at luminosities greater than 10$^{39}$\lum\,with an assumption of isotropic radiation. This critical value is obtained from the Eddington limit (also known as Eddington luminosity) for a 10\,M$_\sun$ object, which sets the maximum luminosity permitted for the balance between its radiation pressure and gravity \citep{1986rpa..book.....R,2004cosp...35.1354M,2018acps.confE..38A}. As the inferred luminosities are so high, it is generally believed that ULXs are accreting binary systems with the primaries being compact objects, such as stellar-mass black holes (SBHs) with super-Eddington accretion or beamed emission, intermediate-mass black holes (IMBHs) \citep{2020ApJ...889...71B}, or neutron stars \citep{2017ARA&A..55..303K}.

Since the first ULX discovered by the Einstein Observatory \citep{1989ARA&A..27...87F}, the studies of ULXs have been active in high-energy astrophysics. Observationally, there were many X-ray survey campaigns that aim to search for new ULXs in nearby galaxies. Some of the most recent examples are the surveys with spectral and temporal analysis of ULXs in NGC~2276, NGC~891 and NGC~1316 \citep{2021AcA....71..261S,2021Univ....8...18E,10.1093/mnras/stab943}. In addition, detailed spectral and temporal analyses for individual systems also give insights into the emission mechanisms in ULXs (e.g., NGC 55~ULX1, M82 ULX, and NGC 5408 X-1; \citealt{2022MNRAS.509.5166J,Kaaret:2006wy,2009ApJ...706L.210S}). Currently, it is known that some ULXs are neutron star binaries and many of these systems likely have super-orbital periods on timescales of 10--100~days (e.g., M82~X-2, NGC 5907 ULX1, NGC~7793 P1; \citealt{2016MNRAS.461.4395K,2016ApJ...827L..13W,2017ApJ...835L...9H,2022ApJ...924...65L}). For black hole scenarios, stellar-mass BHs and IMBHs are both popular candidates of ULXs, and some recent examples include NGC 7793 P9 that has been suggested to be a stellar-mass BH \citep{2018ApJ...864...64H} and M51 ULX-7 a IMBH \citep{10.1093/mnras/stv2945}.

Besides the magnificent powers, some ULXs exhibited interesting periodicities in X-rays that could shed light on their physical natures. As mentioned eariler, super-orbital periods (i.e., $>$ 10 days) can be seen in many ULX pulsars and also in some black hole candidate systems \citep{2015MNRAS.454.1644L}. Periodicities on short time-scales (i.e., $<$ 10 sec) can be seen from pulsar binaries due to the spin of the neutron stars \citep{2022arXiv220104401E}. On hourly to daily time-scales (i.e., $\gtrsim$ 1 day), the periodicities could be the orbital periods of compact binaries \citep{2005ApJ...621L..17L}. In this context, the long-term periodicities would provide information on the orbital parameters of the binaries, which are crucial to constrain their accretion states and the masses of the compact stars \citep{2017ARA&A..55..303K}.

In this paper, we present a \textit{Chandra} study for the bright X-ray point sources in the spiral galaxy NGC~1559. The original purpose of the \textit{Chandra} observation (ObsID: 16745; PI: Kwan-Lok Li) is to verify whether the stellar remnant of the historical supernova Type II-L SN~1986L \citep{1999A&AS..139..531B} becomes ultraluminous in X-rays. While the \textit{Chandra} data were previously analyzed in \cite{8905968}, the absence of temporal analyses for the whole X-ray population and the lack of spectral analyses for the bright non-ULX sources motivate this independent investigation. Some other additional findings, e.g., 9 new X-ray sources that include an ULX are recovered, are also presented in this piece of work.

The paper is organized as follows. In Section \ref{sec:NGC}, we present a brief introduction of NGC 1559. Section \ref{sec:observation} shows the detailed information of \textit{Chandra} observation (ObsID: 16745). Data reduction and analysis processes are presented in Section \ref{sec:data}, and the results in Section \ref{sec:results}. Discussions of the X-ray properties of the bright X-ray sources are presented in Section \ref{sec:discussions}. Finally, we summarize the findings of the X-ray population study in Section \ref{sec:cite}.

\section{NGC~1559}\label{sec:NGC}

NGC~1559 is a barred spiral galaxy that hosts at least four supernovae in the last 40 years \citep{10.1093/mnras/stv857}, including the Type II SN~1984J, Type II-L SN~1986L \citep{1999A&AS..139..531B}, Type Ia SN~2005df \citep{2005CBET..193....1W}, and Type II-P SN~2009ib \citep{2009CBET.2099....1P}. Among these, SN~1986L was the main target of the \textit{Chandra} observation. 

The center of the galaxy is located at $\alpha\mathrm{(J2000)}=04^\mathrm{h}17^\mathrm{m}35\fs77$, $\delta\mathrm{(J2000)}=-62\arcdeg 47\arcmin 1\farcs2$ with a positional uncertainty of 1\farcs25, and its isophotal radius at a surface brightness of 27.0 B-mag\,arcsec$^{-2}$ is 2\farcm7 \citep{2006AJ....131.1163S,1989spce.book.....L}. The Galactic absorption of NGC~1559 is \nh\ $=1.98\times10^{20}$\cm\ \citep{2016A&A...594A.116H}. By using the Tully–Fisher relation (TFR), the distance of NGC 1559 was found to be $12.60$~Mpc (1 arcsec = 61.09 pc; \citealt{2013AJ....146...86T}). Unless otherwise mentioned, we made use of the above spatial and distance information for the \textit{Chandra} analyses as well as the relevant. 

From the visual appearance, in the Hubble sequence's morphological classification, it is a type SB(s)cd galaxy \citep{1991rc3..book.....D}, whose spiral arms are very loose, even fragmentary or discontinuous, and most of the luminosity comes from the arms rather than the bulge \citep{1959HDP....53..275D}. Despite the fragmented morphology, the spiral arms of NGC~1559 are massive with high star formation rate (SFR) \citep{2002A&A...391...83B}. Based on the fact that the distribution of high SFR concentrates on the spiral arms, resent literatures suggest that NGC~1559 may not have an active galactic nucleus (AGN; \citealt{8905968}).

\section{X-ray Observations} \label{sec:observation}

%\subsection{\textit{Chandra} observation}
The \textit{Chandra} X-ray Observatory (CXO) observed NGC~1559 using the Advanced CCD Imaging Spectrometer (ACIS) on June 19, 2016 (ObsID: 16745). The ACIS-S mode was used and the target galaxy was in the field of view of the ACIS S3 back-illuminated chip, which is the most sensitive one compared to the other ACIS chips. The effective exposure time is around 45.4~ks.

Besides the \textit{Chandra} observation, \textit{XMM-Newton} (14.4~ks; ObsID 0403070401) and \textit{Swift} also observed NGC~1559 in 2007 and 2005--2016 (56 observations with a total exposure of 103~ks), respectively. The limited angular resolutions of \textit{XMM-Newton} and \textit{Swift}/XRT prohibit meaningful spatial analyses of a crowded field like NGC~1559 (i.e., more than 30 X-ray sources in a circular region of radius 2\farcm7 plus the diffuse emission of the host galaxy). Nevertheless, we have cross-checked the source list detected by \textit{Chandra} with the 4XMM-DR11 \citep{2020A&A...641A.136W} and 2SXPS \citep{2020ApJS..247...54E} catalogs to check the long-term variabilities. In addition, we selected two X-ray sources that are relatively bright and isolated (i.e., X-1, X-17, and X-23) and another one that exhibited a possible $\sim7500\,$s periodicity (i.e., X-24) for further analyses. The results are presented in Section \ref{sec:results}.

\section{Data reduction} \label{sec:data}

We downloaded the \textit{Chandra} observation (ObsID: 16745) from the \textit{Chandra} Data Archive (CDA) and performed the data reprocessing to create the level 2 event files using the task, \texttt{CHANDRA REPRO}, of the \textit{Chandra} Interactive Analysis of Observation software (\texttt{CIAO}; version 4.13; with CALDB version 4.9.4). For the ACIS datasets, we focused on the energy band of 0.3--7~keV. We adopt the 2\farcm7 radius of the isophote at 27.0 B-mag\,arcsec$^{-2}$ \citep{1989spce.book.....L}, and assume that sources within this isophotal radius belong to NGC~1559 and sources outside are background (or foreground).

The whole X-ray analysis process can be divided into three parts. The first is the detection of X-ray sources. The second part is the spectral fitting analysis. An absorbed (both Galactic and intrinsic \nh\ components are considered) power-law spectral model is generally assumed, but a thermal model was also considered for some relatively soft X-ray sources. The final part is the time-domain analysis of the detected ULXs. X-ray light curves of the ULXs were extracted and examined to check the variabilities and see if short-term periodicities can be found.

\subsection{Source detection}
We used the software, \texttt{CIAO} (version 4.13; with CALDB version 4.9.4), developed by CXO to conduct the above processes. We used the task \texttt{WAVDETECT}, which convolutes the input X-ray image with a ``Mexican Hat'' filter with optimized scales as the detection method for point sources. Before running \texttt{WAVDETECT}, we first created an exposure-corrected image by \texttt{FLUXIMAGE}. This step was necessary because of the nonuniform sensitivities across the ACIS chips. Despite the fact that the deviations around the aimpoint of the observation (i.e., within the isophotal radius of NGC~1559) are usually small, extragalactic X-ray sources are often so faint that even tiny deviations cannot be ignored.

The major input parameters of \texttt{FLUXIMAGE} are \texttt{binsize}, \texttt{bands} and \texttt{psfecf}. The \texttt{binsize} and \texttt{psfecf} were set to 1 and 0.393, respectively, as suggested by the CXO analysis thread\footnote{\url{https://cxc.cfa.harvard.edu/ciao/threads/wavdetect/}}. The \texttt{band} was set to 0.3--7~keV as mentioned earlier. The exposure-corrected image from \texttt{FLUXIMAGE} was then fed to \texttt{WAVDETECT}. The wavelet scales were set to 1, 2, 4, 8 and 16 pixels, which are commonly adopted for \textit{Chandra}-ACIS (e.g., \citep{2011MNRAS.414.1329C}). All other parameters were kept default. \texttt{WAVDETECT} detected 32 X-ray sources. We visually inspected the result to remove invalid sources (e.g., noise detected by \texttt{WAVDETECT}) from the source list and also check for any obvious X-ray sources missed by \texttt{WAVDETECT}.

\subsection{Spectral analysis}
The next step is to extract and analyze spectra of these sources using \texttt{XSPEC} in the \texttt{HEASoft} package (version 6.28), developed by NASA. The main goal is to fit spectral models to the data and assess physical properties of the sources. We used the \texttt{CIAO} task, \texttt{SPECEXTRACT}, to extract spectra from the level 2 event file produced by \texttt{CHANDRA\_REPRO}. The regions of the detected X-ray sources are estimated by the task \texttt{WAVDETECT}. The radii of these sources are about 1--3 arcsec, depending on the X-ray brightness. For the background, it is hard to assign a dedicated background region for every X-ray source as the field is crowded and full of diffuse X-ray emission. Thus, we chose one source-free region to represent the background of NGC 1559 and the coordinates of the background region are $\alpha\mathrm{(J2000)}=04^\mathrm{h}17^\mathrm{m}25\fs80$, $\delta\mathrm{(J2000)}=-62\arcdeg 47\arcmin 55\farcs43$ with a 5\arcsec\ radius. We also turned on the \texttt{correctpsf} feature to allow aperture correction, which compensates the lost source count outside the source region of a source. After the extraction, we used the \texttt{HEASoft} task \texttt{GRPPHA} to bin the X-ray spectra. 

We used a power-law model with two absorption components as the spectral model. The first one referring to the Galactic absorption is fixed at \nh\ $=1.98\times10^{20}$\cm\ \citep{2016A&A...594A.116H}. The second component could be adjusted, corresponding to the intrinsic absorption of the host galaxy and/or the X-ray emitting systems. 

Since the faintest and the brightest sources in the \texttt{WAVDETECT} list can differ by a factor of $>10$, we divide the sources into three groups according to the numbers of source counts detected (i.e., Group A: more than 100 counts; Group B: between 35 and 100 counts; Group C: less than 35 counts), each of which has a slightly different fitting strategy. In Group A, we used the generic power-law model mentioned above with the intrinsic \nh\ and the photon index as free parameters. As the sources in the group have sufficient numbers of counts, we grouped at least 20 spectral counts into 1 bin for this group to increase the signal-to-noise ratio per bin, and $\chi^2$ was adopted as the fitting statistic. For Group B, the same spectral model as Group A was applied, but the binning factor was set to 1 to avoid low degrees of freedom and \textit{C-statistic} ($C$; \citealt{1979ApJ...228..939C}) was therefore used in the fitting process. The strategy of Group C is basically the same as that of Group B. However, considering the limited number of counts of the spectra, we further simplified the spectral model by freezing the intrinsic absorption to \nh\ $=1.7\times10^{21}$\cm, the smallest best-fit \nh\ obtained in Group A, which can be regarded as the minimum intrinsic \nh\ of the host galaxy as observed from Earth. In this sense, the inferred X-ray luminosities are better seen as lower limits. 
On the other hand, the X-ray sources that are not located in the galactic disk regions might have lower intrinsic \nh\ values, and the luminosities could be slightly overestimated.
Based on the criterion of categories, the number of the detected sources in Group A, B and C are 9, 7 and 17, respectively.

We emphasize that the estimation method is not ideal, but it is already the best way given the low photon statistics. Based on the spectral parameters of the models, like the \nh\ values and photon index ($\Gamma$), we calculate the unabsorbed X-ray fluxes ($F_x$) and the X-ray luminosities ($L_x$) of the sources (12.6~Mpc was assumed). We then classified whether an X-ray source is a ULX with the criterion of $L_x>10^{39}$\lum.

\subsubsection{Hardness ratios}
For the reason that many of the detected sources have low photon statistics, we also computed the X-ray hardness ratios of the X-ray sources to show their X-ray colors. The soft X-ray color/hardness is defined as 
   \begin{equation}
      {\mathit{HR_1}=\frac{\mathit{M}-\mathit{S}}{\mathit{M}+\mathit{S}}\text{,}}
   \end{equation}
while the hard color (hardness) is 
   \begin{equation}
      {\mathit{HR_2}=\frac{\mathit{H}-\mathit{M}}{\mathit{H}+\mathit{M}}\text{,}}
   \end{equation}
where $S$, $M$ and $H$ are the net counts in 0.3--1 keV, 1--2 keV, and 2--7 keV, respectively.

\subsection{Light curves and search for periodicities}
The light curves of the X-ray sources were extracted by \texttt{DMEXTRACT} in \texttt{CIAO}. The source/background regions and the corrected exposure images are the same as those used in the aforementioned processes. For the uncertainty estimates, Gaussian noise (i.e., noise $=\sqrt{N}$, where $N$ is the count number) was assumed for the source counts, while the Gehrels approximation (i.e., noise $=1+\sqrt{N+0.75}$; \citealt{1986ApJ...303..336G}) was employed for the low-count background. In addition, the "bin time'' parameter was set to 10 seconds, which is small enough to keep the temporal information required for a periodicity search, in which the light curves will be further binned.

Visual inspection was first used to look for any obvious periodic signals in the light curves (1000-s re-binned), and we did not find any signal. It is somewhat expected as the X-ray sources are rather faint in this study (i.e., the numbers of source counts are all less than $1000$). 

For a more careful and systematic search, we examined the periodograms produced by the \texttt{HEASoft} task, \texttt{EFSEARCH}, which phases the light curve on a given range of periods using epoch folding and computes the $\chi^2$ statistic for each phased light curve with an assumption of a flat light curve model. The number of phase bins was set as 10 for a balance between the signal-to-noise ratio per bin and the timing resolution. We searched all the sources in the range of period from 1000s to 22000s with a resolution of 200s and focused on the sources whose $\chi^2$ have an apparent difference between maximum and minimum, which implies the sources have a periodicity potentially. Afterward, the candidate periodic sources will be searched in the same range of period with a higher resolution of 10s.

%--------------------------------------------------------------------
   \begin{figure*} [htb]
	\centering
	\includegraphics[width=\hsize]{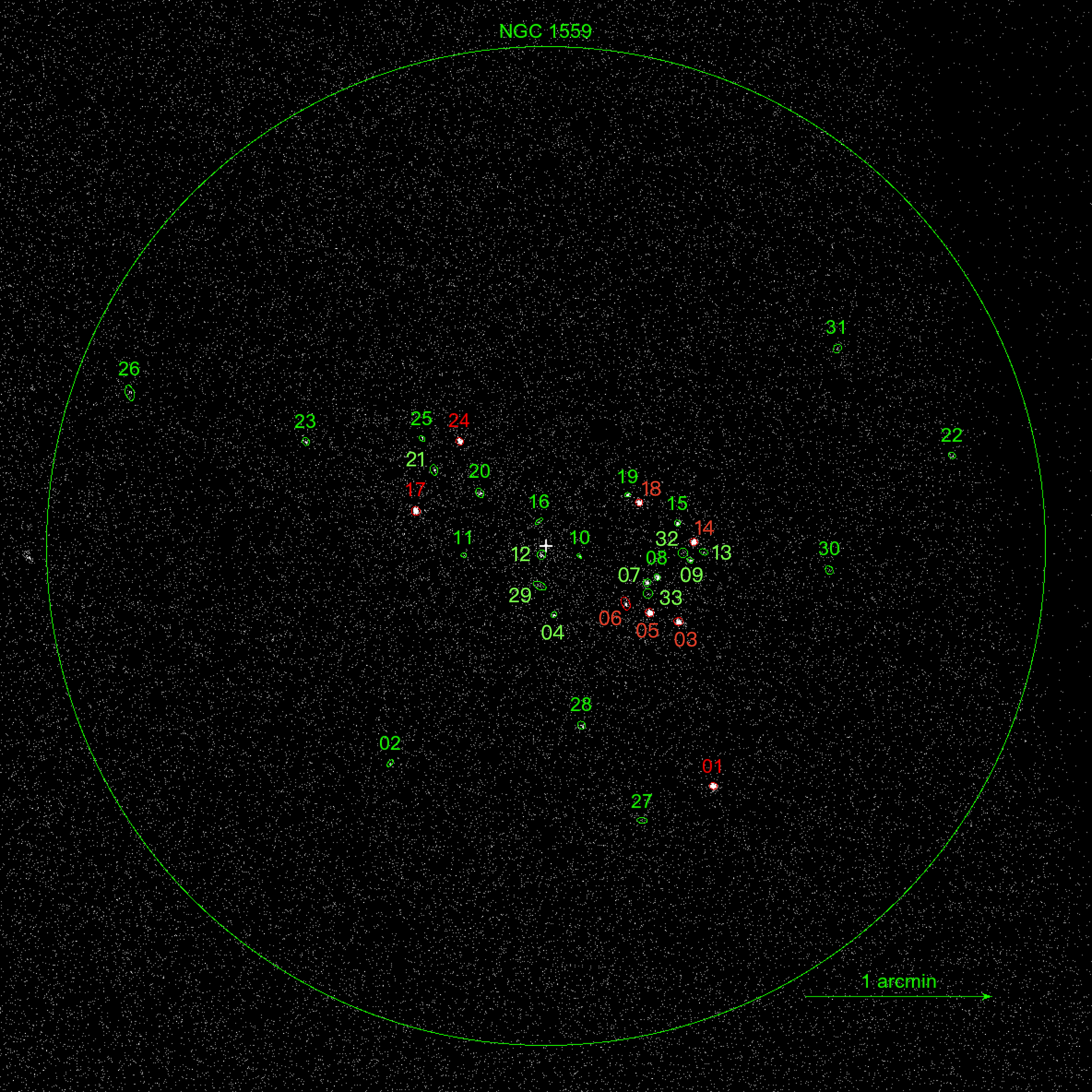}
	\caption{The population of X-ray sources in NGC~1559. The \textit{Chandra} X-ray image (0.3--7~keV) of NGC~1559. The peripheral circle is the boundary of NGC 1559 whose isophotal radius is 2\farcm7. The coordinate of center marked by the white cross-hairs is at $\alpha\mathrm{(J2000)}=04^\mathrm{h}17^\mathrm{m}35\fs77$, $\delta\mathrm{(J2000)}=-62\arcdeg 47\arcmin 1\farcs2$. The detected X-ray sources are indicated by green and red elliptical regions, which are optimized to "Mexican Hat" wavelet functions estimated with different sizes by \texttt{WAVDETECT}, except for X-32 and X-33 which are added manually. The red regions mark the ULXs. The given source IDs near the ellipse are the same as those used in Table \ref{tab:1}.}
	\label{fig:population}
   \end{figure*}
%--------------------------------------------------------------------
   \begin{figure*} [p]
	\begin{center}
	\includegraphics[width=\hsize]{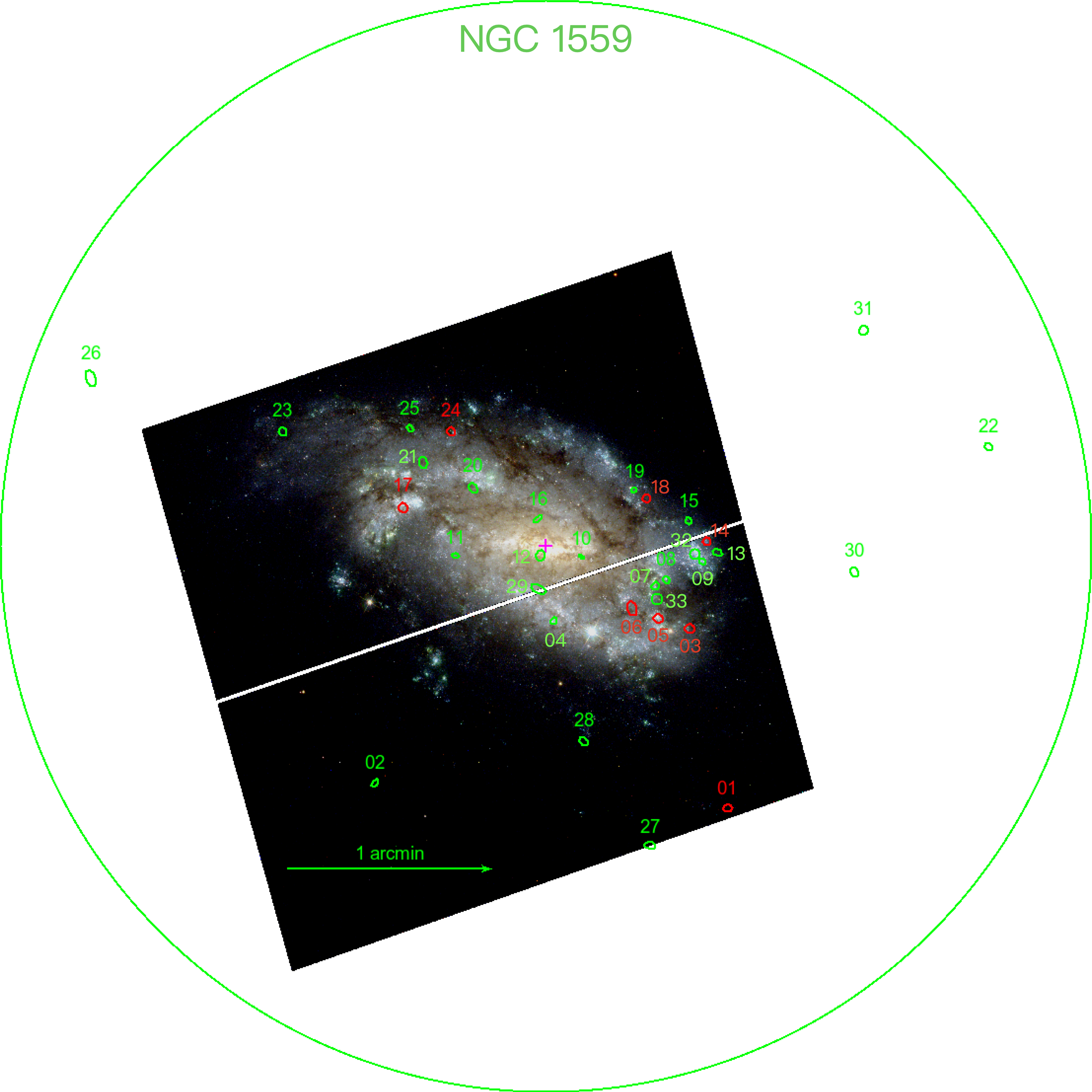}
	\caption{The optical image of NGC~1559 taken by HST/WFC3 obtained from the HLA (F814W: red; F606W: green; F438W: blue). We place the positions of the X-ray sources detected by \textit{Chandra} on the image, which are the same as the ones shown in Figure \ref{fig:population}. The field of view of the HST image is smaller than that of the \textit{Chandra} image. Thus, some of sources were missed by HST.}
	\label{fig:Hl}
	\end{center}
   \end{figure*}
%----------------------------------------------------------------- 

\subsection{\textit{Swift}/XRT and \textit{XMM-Newton}/EPIC}

As we have mentioned in Section \ref{sec:observation}, X-1, X-17, X-23, and X-24 were selected for further analyses with the \textit{Swift}/XRT and \textit{XMM-Newton}/EPIC observations. For the \textit{Swift}/XRT observations, we used \texttt{HEASoft} together with \textit{Swift}/XRT data products generator\footnote{\url{https://www.swift.ac.uk/user_objects/index.php}} \citep{2007A&A...469..379E,2009MNRAS.397.1177E} to reduce the data and extract the X-ray light curves and spectra (0.3--10~keV). The observations taken in the Windowed Timing (WT) mode were not used, given that the NGC~1559 field is very crowded. The \textit{Chandra} coordinates were directly used as the inputs. Circular regions of $10\arcsec$ radius were adopted for source extraction, except for X-1 that is rather isolated and thus radius 20\arcsec\ was used for collecting more events. The background extraction region was a source-free circular region of $50\arcsec$ radius.

For the \textit{XMM-Newton} observation, the data was downloaded from the HEASARC data server and the X-ray light curve/spectra (0.2--10~keV) were extracted by the task \texttt{xmmextractor} from \texttt{SAS} (version: 19.1.0). Similar to the process for \textit{Swift}/XRT, the \textit{Chandra} positions were used without re-centering the sources. The source regions are circular with 10\arcsec\ radius and the background was extracted with a source-free circular region of $50\arcsec$ radius. The task also performed background filtering to minimise the effects from the external soft proton flares on the X-ray data. The pn data suffers heavily from the flaring background (only 3.2~ks left after the background filtering), while the effects on the MOS-1 and MOS-2 observations are minor (11.8~ks and 13.4~ks, respectively). Therefore, we focused only on the MOS-1/2 data in this work.

For both the \textit{Swift}/XRT and \textit{XMM-Newton} MOS-1/2 spectra, the spectra were binned according to the data quality. A binning factor of minimum 20 counts per bin was used, if the data quality is allowed. In this case, $\chi^2$ was employed as the fitting statistic. Otherwise, minimum 1 counts per bin was used, and \textit{C-statistic} (\citealt{1979ApJ...228..939C}) was employed.

\section{Results} \label{sec:results}

\subsection{Source detection}
Using the process described in the previous section, \texttt{WAVDETECT} detected 32 X-ray sources in NGC~1559. We examined the source regions extracted by \texttt{WAVDETECT} in the \textit{Chandra} image to check the result. If the detection significance is greater than 3, the source is considered significant. All the sources look good, except for one whose significance is just 1.38 $\sigma$. Visually, the source has a region of zero area (i.e., an ellipses with one of the axes equal to zero) and also no apparent net count is seen at the position. Thus, we concluded that the source is unreal and was rejected from the final source list.

In addition to these 31 \texttt{WAVDETECT} sources, we visually found two extra sources that have net counts of 17 and 12, respectively. In spite of the null detections from \texttt{WAVDETECT}, the two sources are both $>3\sigma$ above the fluctuation of the background, and therefore, they were also added in the final list. In total, 33 sources were detected, and they are listed in Table \ref{tab:1} and shown in Figure \ref{fig:population}, with the source IDs assigned by \texttt{WAVDETECT} (X-32 and X-33 are the two missed by \texttt{WAVDETECT}). 
These 33 X-ray sources include all the 24 sources reported in \cite{8905968}.

It is also worth mentioning that none of them are coincident with the historical supernovae in NGC 1559. Besides the X-ray observations, Hubble Space Telescope (HST) observed this galaxy with the Wide Field Camera 3 (WFC3) in 2015. We downloaded the images from the Hubble Legacy Archive (HLA) and overlaid all X-ray sources on it to show the spatial distribution of the X-ray sources. (Figure \ref{fig:Hl}).

%--------------------------------------------------------------------

\subsection{X-ray spectra and luminosities}
The best-fit parameters (and the corresponding 90\% uncertainties) for the absorbed power-law model and the hardness ratios of the 33 sources are all listed in Table \ref{tab:1}. All the fluxes and luminosities listed are absorption-corrected, and the 90\% uncertainties were computed using the \texttt{XSPEC} command \texttt{error}. In Group A, except for X-18, all the best-fit photon index in the group are around $\Gamma\approx$ 1--2, indicating that they are probably non-thermal X-ray systems. In the case of ULX X-18, the photon index is $\Gamma=2.34^{+0.38}_{-0.34}$. Although it is not particularly soft, a thermal emission origin is possible and therefore we added an additional blackbody component to the simple power-law model. The best-fit blackbody temperature is 0.34$\pm_{0.04}^{0.06}$~keV with a normalization of $R^2_{km}/D^2_{10}=0.33\pm_{0.20}^{0.22}$, where $R_{km}$ is the source radius in km and $D_{10}$ is the distance to the source in units of 10 kpc. The $\chi^2$ is 25.73/17. The likelihood ratio test was used to test the significance of the improvement. The chance probability is 39\%, indicating the improvement is insignificant. 

In Groups B and C, there are also a few relatively soft X-ray sources that have best-fit photon indices above $\Gamma=2$. However, given that the numbers of spectral counts are low in these groups (i.e., less than 100 counts) and the uncertainties are large, we did not further consider any advanced models for them.

We also cross-checked the X-ray fluxes of the 24 sources reported in (\citealt{8905968}; which were estimated by WebPIMMS assuming the \nh\ values obtained from \texttt{HEASoft} and a fixed photon index of 1.9) with ours. The results are mostly consistent, except for the ones with high intrinsic \nh\ (e.g., X-6 and X-17) and photon indices very different from 1.9 (e.g., X-10 and X-25).

\begin{figure}[t]
      \begin{center}
      \subfigure[The spectrum of X-24]{\label{S-24}\includegraphics[width=\hsize]{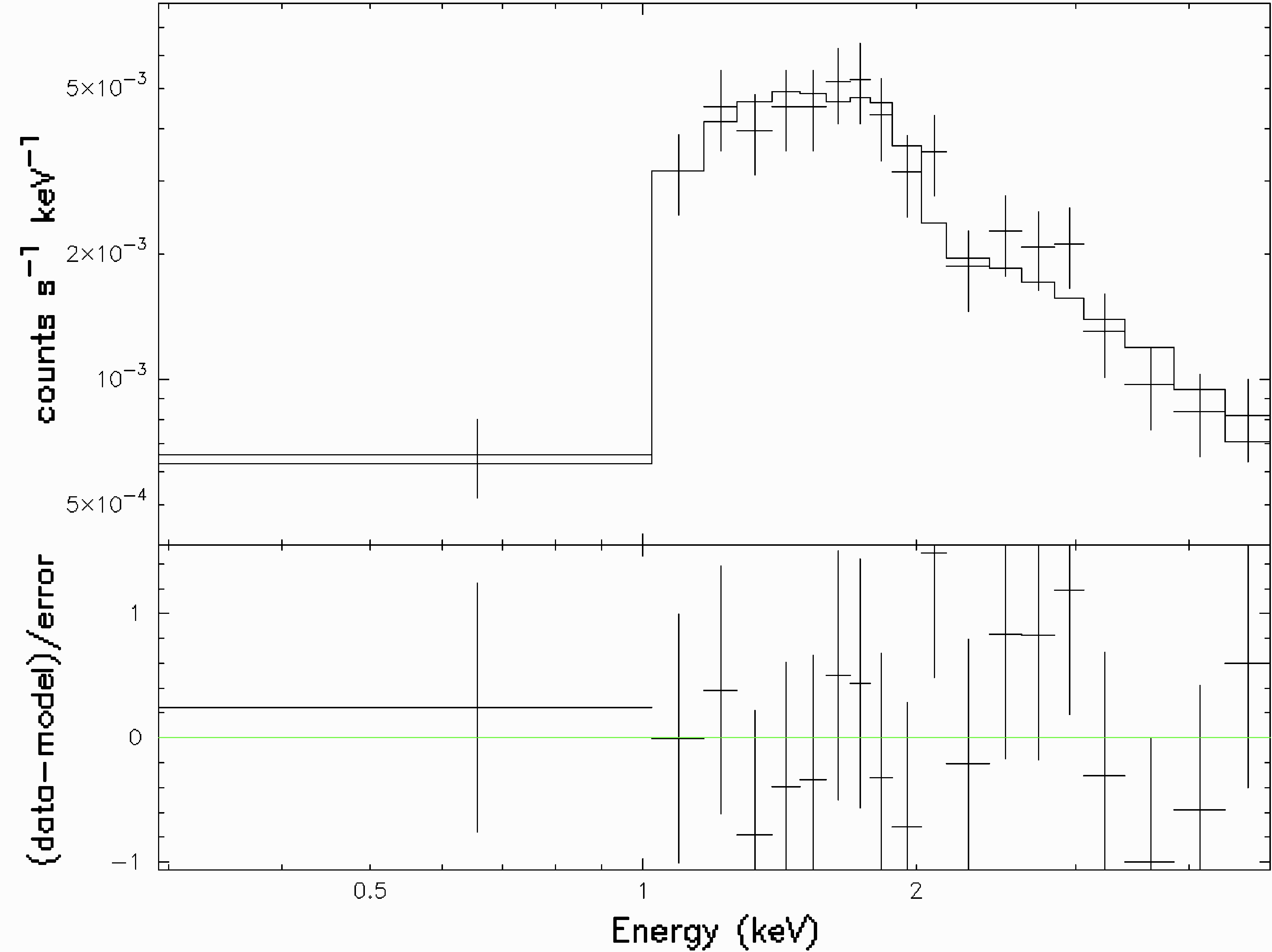}}
      \subfigure[The spectrum of X-06]{\label{S-06}\includegraphics[width=\hsize]{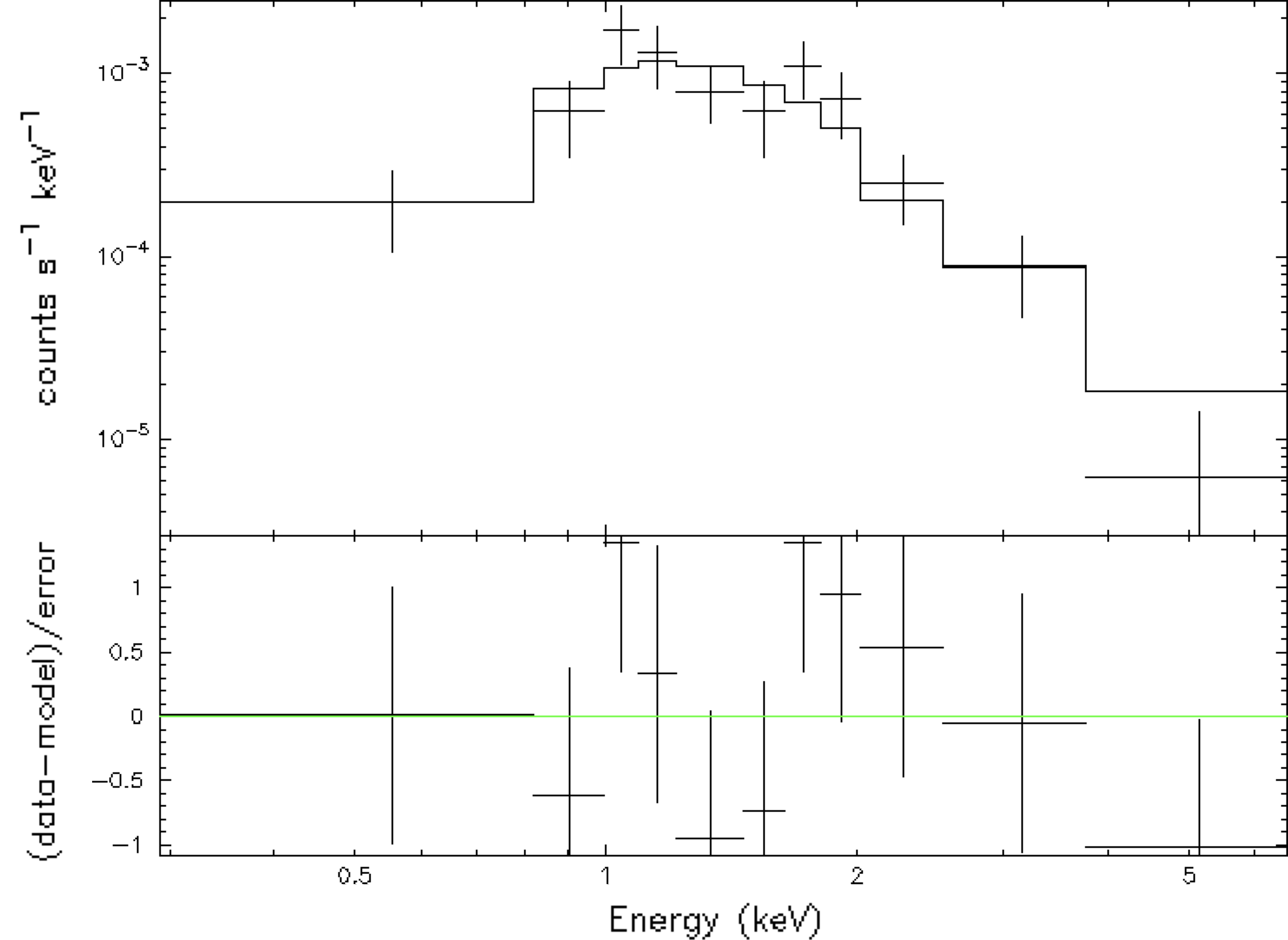}}
      \caption{The spectra of X-24 and X-06. The two diagrams present the different qualities of data by spectra. The diagram (a), the spectrum of X-24, represents a case of good quality. Meanwhile, X-24 is also the only periodic X-ray binary in NGC 1559. The diagram (b), the spectrum of X-06, represents a contrasting example of bad quality. The error bars of X-06 are apparently very large.}
      \label{fig:spectra}
      \end{center}
   \end{figure}

Based on the spectral parameters and the assumed distance to the host galaxy, we computed the X-ray luminosities, with which the ULX classification was made. 8 ULXs were identified based on the criteria of $L_x>10^{39}$\lum. 7 of them belong to Group A (with 100 or more source counts) and only one is in Group B (X-6; with 35--100 source counts). Despite the high X-ray luminosity, almost half of the ULX spectra do not provide well constrained spectral parameters owing to the insufficient photon statistics. In Figure \ref{fig:spectra}, the X-ray spectra of X-06 (among the worst) and X-24 (among the best) are shown for reference.
   
   %-------------------------------------------------------------------- Two column figure (place early!)
   \begin{table*} [p]
    \scriptsize
    \caption{The detected X-ray sources including the detail properties in NGC~1559}             % title of Table
        \label{tab:1}
    \tabcolsep=3pt
    \renewcommand\arraystretch{1.6}
    \vspace{-0.4cm}
\begin{center}    
    \begin{tabular}{c c c c c c c c c c c}       % centered columns (4 columns)
    \hline\hline                                      % inserts double horizontal lines

     \multirow{2}{*}{ID} & \multirow{2}{*}{R.A.} & \multirow{2}{*}{Dec.} & \nh\ & \multirow{2}{*}{$\Gamma$} & log$_{10}$F & $L_x$ & \multirow{2}{*}{$\rm{HR_1}$} & \multirow{2}{*}{$\rm{HR_2}$} & Statistic/dof & \multirow{2}{*}{Group}  \\
    & & & ($10^{22}$\cm) & & (\flux) & ($10^{39}$\lum) & & & ($\chi^2$ or $C$) & \\ 
    
    \hline                                               % inserts single horizontal line

1   & 4:17:27.98 & -62:48:17.94 & 0.24$^{+0.13}_{-0.11}$ & 1.81$^{+0.26}_{-0.24}$ & -12.66$^{+0.07}_{-0.06}$ & 4.15$_{-0.54}^{+0.73}$  & 0.69$\pm$0.04 & -0.11$\pm$0.04 & 24.8/25 & A \\
2   & 4:17:43.06 & -62:48:10.59 & [0.17]                             & 1.73$^{+0.93}_{-0.91}$ & -14.26$^{+0.18}_{-0.22}$ & 0.10$_{-0.04}^{+0.05}$  & 0.50$\pm$0.31 & 0.00$\pm$0.30  & 8.2/12 & C \\
3   & 4:17:29.60 & -62:47:25.31 & 0.37$^{+0.23}_{-0.18}$ & 1.47$^{+0.38}_{-0.34}$ & -12.87$^{+0.09}_{-0.07}$ & 2.57$_{-0.39}^{+0.58}$  & 0.79$\pm$0.05 & 0.04$\pm$0.06  & 4.3/13 & A \\
4   & 4:17:35.42 & -62:47:23.07 & [0.17]                             & 2.48$^{+0.67}_{-0.65}$ & -13.89$^{+0.19}_{-0.16}$ & 0.25$_{-0.08}^{+0.14}$  & 0.48$\pm$0.18 & -0.36$\pm$0.19 & 18.3/28 & C \\
5   & 4:17:30.95 & -62:47:22.38 & 0.41$^{+0.18}_{-0.14}$ & 1.81$^{+0.27}_{-0.24}$ & -12.56$^{+0.09}_{-0.07}$ & 5.25$_{-0.79}^{+1.19}$  & 0.74$\pm$0.04 & -0.04$\pm$0.04 & 29.4/28 & A \\
6   & 4:17:32.07 & -62:47:19.50 & 0.37$^{+0.35}_{-0.27}$ & 3.16$^{+1.04}_{-0.86}$ & -13.25$^{+0.64}_{-0.40}$ & 1.06$_{-0.64}^{+3.56}$  & 0.63$\pm$0.11 & -0.55$\pm$0.11 & 36.9/48 & B \\
7   & 4:17:31.07 & -62:47:12.83 & 0.37$^{+0.61}_{-0.37}$ & 1.31$^{+0.88}_{-0.73}$ & -13.36$^{+0.23}_{-0.12}$ & 0.83$_{-0.20}^{+0.57}$  & 0.85$\pm$0.08 & 0.05$\pm$0.10  & 0.8/2 & A \\
8   & 4:17:30.60 & -62:47:11.08 & 0.17$^{+0.46}_{-0.17}$ & 1.50$^{+1.10}_{-0.71}$ & -13.38$^{+0.27}_{-0.12}$ & 0.79$_{-0.19}^{+0.69}$  & 0.59$\pm$0.10 & -0.16$\pm$0.10 & 0.0/2 & A \\
9   & 4:17:29.06 & -62:47:05.69 & 0.11$^{+0.35}_{-0.11}$ & 1.79$^{+0.91}_{-0.64}$ & -13.74$^{+0.30}_{-0.15}$ & 0.35$_{-0.10}^{+0.34}$  & 0.59$\pm$0.14 & -0.23$\pm$0.15 & 32.7/42 & B \\
10 & 4:17:34.22 & -62:47:04.35 & [0.17]                             & 0.93$^{+1.09}_{-1.11}$ & -13.73$^{+0.30}_{-0.28}$ & 0.35$_{-0.17}^{+0.35}$  & 1.00$\pm$0.01 & 0.20$\pm$0.31  & 10.5/8 & C \\
11 & 4:17:39.61 & -62:47:04.01 & [0.17]                             & 1.15$^{+0.97}_{-0.98}$ & -14.17$^{+0.25}_{-0.24}$ & 0.13$_{-0.05}^{+0.10}$  & 1.00$\pm$0.01 & -0.17$\pm$0.29 & 9.0/10 & C \\
12 & 4:17:35.99 & -62:47:03.92 & 0.12$^{+0.36}_{-0.12}$ & 1.89$^{+0.97}_{-0.69}$ & -13.82$^{+0.35}_{-0.19}$ & 0.29$_{-0.10}^{+0.36}$  & 0.63$\pm$0.14 & -0.27$\pm$0.15 & 37.8/38 & B \\
13 & 4:17:28.42 & -62:47:03.04 & [0.17]                             & 1.45$^{+0.69}_{-0.66}$ & -14.06$^{+0.15}_{-0.16}$ & 0.16$_{-0.05}^{+0.07}$  & 0.74$\pm$0.18 & -0.13$\pm$0.21 & 14.2/23 & C \\
14 & 4:17:28.89 & -62:46:59.79 & 0.18$^{+0.14}_{-0.11}$ & 1.67$^{+0.28}_{-0.25}$ & -12.75$^{+0.07}_{-0.06}$ & 3.38$_{-0.44}^{+0.59}$  & 0.64$\pm$0.04 & -0.10$\pm$0.05 & 20.9/21 & A \\
15 & 4:17:29.65 & -62:46:53.75 & 0.07$^{+0.25}_{-0.07}$ & 2.50$^{+0.96}_{-0.59}$ & -13.63$^{+0.44}_{-0.17}$ & 0.44$_{-0.14}^{+0.78}$  & 0.51$\pm$0.12 & -0.51$\pm$0.12 & 29.9/46 & B \\
16 & 4:17:36.11 & -62:46:53.21 & [0.17]                             & 2.40$^{+1.58}_{-1.22}$ & -14.40$^{+0.44}_{-0.31}$ & 0.08$_{-0.04}^{+0.13}$  & 0.14$\pm$0.39 & -0.15$\pm$0.39 & 12.5/8 & C \\
17 & 4:17:41.86 & -62:46:49.86 & 1.52$^{+0.58}_{-0.44}$ & 1.85$^{+0.50}_{-0.43}$ & -12.38$^{+0.23}_{-0.13}$ & 7.98$_{-2.06}^{+5.57}$  & 0.96$\pm$0.02 & 0.32$\pm$0.04  & 28.1/22 & A \\
18 & 4:17:31.44 & -62:46:47.20 & 0.23$^{+0.12}_{-0.10}$ & 2.34$^{+0.38}_{-0.34}$ & -12.72$^{+0.14}_{-0.10}$ & 3.64$_{-0.75}^{+1.39}$  & 0.60$\pm$0.04 & -0.33$\pm$0.05 & 25.9/19 & A \\
19 & 4:17:31.99 & -62:46:44.71 & 0.24$^{+0.42}_{-0.23}$ & 1.70$^{+0.86}_{-0.74}$ & -13.56$^{+0.29}_{-0.16}$ & 0.52$_{-0.16}^{+0.49}$  & 0.81$\pm$0.11 & -0.19$\pm$0.14 & 30.8/44 & B \\
20 & 4:17:38.87 & -62:46:44.10 & 0.08$^{+0.50}_{-0.01}$ & 0.97$^{+1.00}_{-0.47}$ & -13.85$^{+0.16}_{-0.15}$ & 0.27$_{-0.08}^{+0.12}$  & 0.62$\pm$0.18 & 0.00$\pm$0.17  & 26.7/32 & B \\
21 & 4:17:41.00 & -62:46:36.68 & [0.17]                             & 0.21$^{+0.70}_{-0.75}$ & -13.97$^{+0.18}_{-0.20}$ & 0.20$_{-0.08}^{+0.11}$  & 0.63$\pm$0.40 & 0.65$\pm$0.17  & 19.5/20 & C \\
22 & 4:17:16.85 & -62:46:32.06 & [0.17]                             & -0.04$^{+0.94}_{-1.06}$ & -14.06$^{+0.26}_{-0.27}$ & 0.17$_{-0.08}^{+0.14}$ & 1.00$\pm$0.10 & 0.73$\pm$0.19  & 10.5/12 & C \\
23 & 4:17:46.99 & -62:46:27.64 & 0.43$^{+0.45}_{-0.34}$ & 2.69$^{+1.12}_{-0.90}$ & -13.41$^{+0.64}_{-0.32}$ & 0.73$_{-0.38}^{+2.47}$  & 0.80$\pm$0.10 & -0.39$\pm$0.13 & 49.3/42 & B \\
24 & 4:17:39.80 & -62:46:27.56 & 0.53$^{+0.26}_{-0.21}$ & 1.70$^{+0.39}_{-0.35}$ & -12.71$^{+0.12}_{-0.08}$ & 3.67$_{-0.62}^{+1.17}$  & 0.83$\pm$0.04 & 0.05$\pm$0.05  & 9.0/16 & A \\
25 & 4:17:41.55 & -62:46:26.67 & [0.17]                             & 0.87$^{+0.97}_{-1.01}$ & -14.09$^{+0.28}_{-0.25}$ & 0.15$_{-0.07}^{+0.14}$  & 0.43$\pm$0.35 & 0.24$\pm$0.28  & 19.3/12 & C \\
26 & 4:17:55.19 & -62:46:11.96 & [0.17]                             & 2.63$^{+1.44}_{-1.13}$ & -14.00$^{+0.41}_{-0.31}$ & 0.19$_{-0.09}^{+0.30}$  & 0.25$\pm$0.26 &-0.19$\pm$0.26 & 24.1/20 & C \\
27 & 4:17:31.30 & -62:48:28.92 & [0.17]                             & -0.25$^{+1.12}_{-1.41}$ & -14.22$^{+0.28}_{-0.32}$ & 0.11$_{-0.06}^{+0.10}$ & 1.00$\pm$0.32 & 0.84$\pm$0.20  & 6.6/8 & C \\
28 & 4:17:34.13 & -62:47:58.44 & [0.17]                             & 1.64$^{+0.97}_{-1.05}$ & -14.27$^{+0.20}_{-0.24}$ & 0.10$_{-0.04}^{+0.06}$  & 0.80$\pm$0.22 & -0.15$\pm$0.27 & 10.7/13 & C \\
29 & 4:17:36.07 & -62:47:13.80 & [0.17]                             & 3.07$^{+1.34}_{-1.18}$ & -14.09$^{+0.45}_{-0.32}$ & 0.16$_{-0.08}^{+0.28}$  & 0.23$\pm$0.28 & -0.36$\pm$0.29 & 17.9/15 & C \\
30 & 4:17:22.57 & -62:47:08.80 & [0.17]                             & -0.56$^{+1.63}_{-0.56}$ & -14.19$^{+0.50}_{-0.45}$ & 0.12$_{-0.08}^{+0.27}$ & 1.00$\pm$0.14 & 0.59$\pm$0.30  & 10.9/7 & C \\
31 & 4:17:22.19 & -62:45:57.92 & [0.17]                             & 1.85$^{+1.42}_{-1.26}$ & -14.53$^{+0.32}_{-0.32}$ & 0.06$_{-0.03}^{+0.06}$  & 1.00$\pm$0.06 & 0.00$\pm$0.34  & 12.8/8 & C \\
32 & 4:17:29.36 & -62:47:03.48 & [0.17]                             & 1.81$^{+0.17}_{-1.07}$ & -14.27$^{+0.22}_{-0.22}$ & 0.10$_{-0.04}^{+0.07}$  & 0.08$\pm$0.32 & 0.08$\pm$0.30  & 20.5/16 & C \\
33 & 4:17:30.99 & -62:47:16.55 & [0.17]                             & 2.68$^{+1.55}_{-1.73}$ & -14.19$^{+0.42}_{-0.29}$ & 0.12$_{-0.06}^{+0.20}$  & 0.85$\pm$0.19 & -0.71$\pm$0.23 & 10.6/10 & C \\

    \hline                                   %inserts single line
    \end{tabular}
    \end{center}

\tablecomments{The fluxes and luminosities are all absorption-corrected. The relative locations of the sources in NGC 1559 are circled with green colors in Fig \ref{fig:population}. The absorption models contain the Galactic absorption and the intrinsic absorption. The Galactic absorptions of all are the same which are fixed at $0.0198 \times 10^{22}\ cm^{-2}$. We do not note them in the table. The intrinsic absorptions are presented in the fourth column. The Group C sources have their intrinsic absorptions fixed at the value, $0.17 \times 10^{22}\ cm^{-2}$. The tenth column shows $\chi ^2$ or $C$. In the Group A, the positions show $\chi ^2$. In the Group B and C, the positions show $C$. The last column, Group, labels their groups.}
    \end{table*}

X-06 is the weakest ULX with $L_x\approx1.06_{-0.64}^{+3.56}\times10^{39}$\lum. It could fall into the non-ULX regime, provided that the true X-ray luminosity is at the lower bound of the 90\% confidence interval.
We also comment that it is the only ULX in Group B, probably because of the soft photon index, and hence a high conversion factor between the observed count rate and the corrected flux.
The most luminous ULX in NGC~1559 is X-17, which has an X-ray luminosity of $L_x\approx7.98_{-2.06}^{+5.57}\times10^{39}$\lum\ in 0.3--7~keV. It was also the only ULX exceeding $10^{40}$\lum\ in the galaxy when considering the upper bound of the error. We note that X-17 has a large intrinsic X-ray absorption of \nh $=1.52^{+0.58}_{-0.44}\times 10^{22}$\cm. Although the X-ray luminosity inferred from spectral fitting generally increases with \nh\ value, the photon index (which is also sensitive to the X-ray absorption) of the ULX is hard (i.e., $\Gamma<2$; Table \ref{tab:1}), suggesting that the high \nh\ is unlikely artificial. In addition, the source counts of X-17 are 525 that is among the highest in the sample. Therefore, we believe that the inferred X-ray luminosity of X-17 is genuine.

%--------------------------------------------------------------------

\subsection{Timing analysis}

\begin{figure*}[p]
     \begin{center}
     \subfigure[X-01]{\label{X-01}\includegraphics[width=0.329\textwidth]{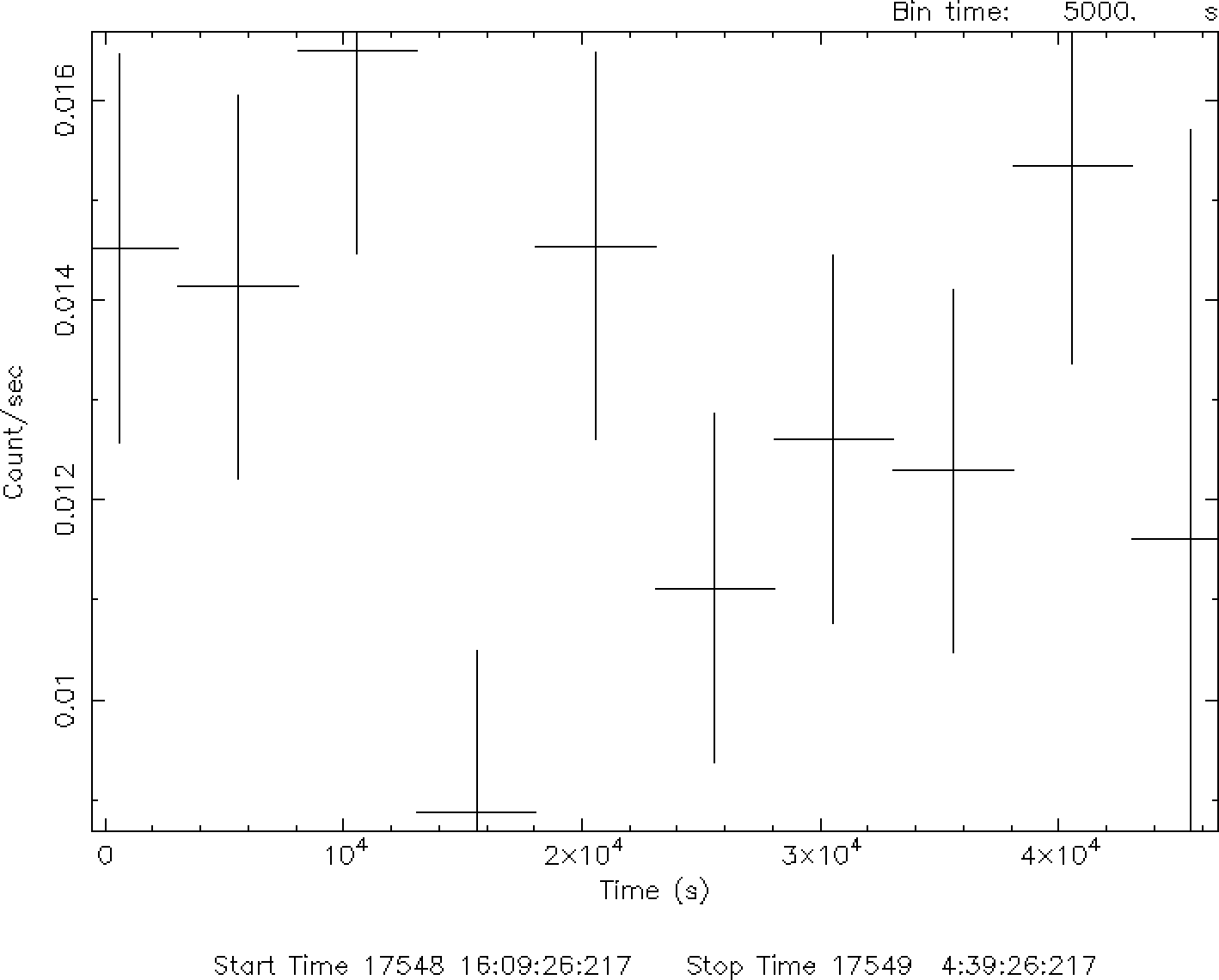}}
     \subfigure[X-03]{\label{X-03}\includegraphics[width=0.329\textwidth]{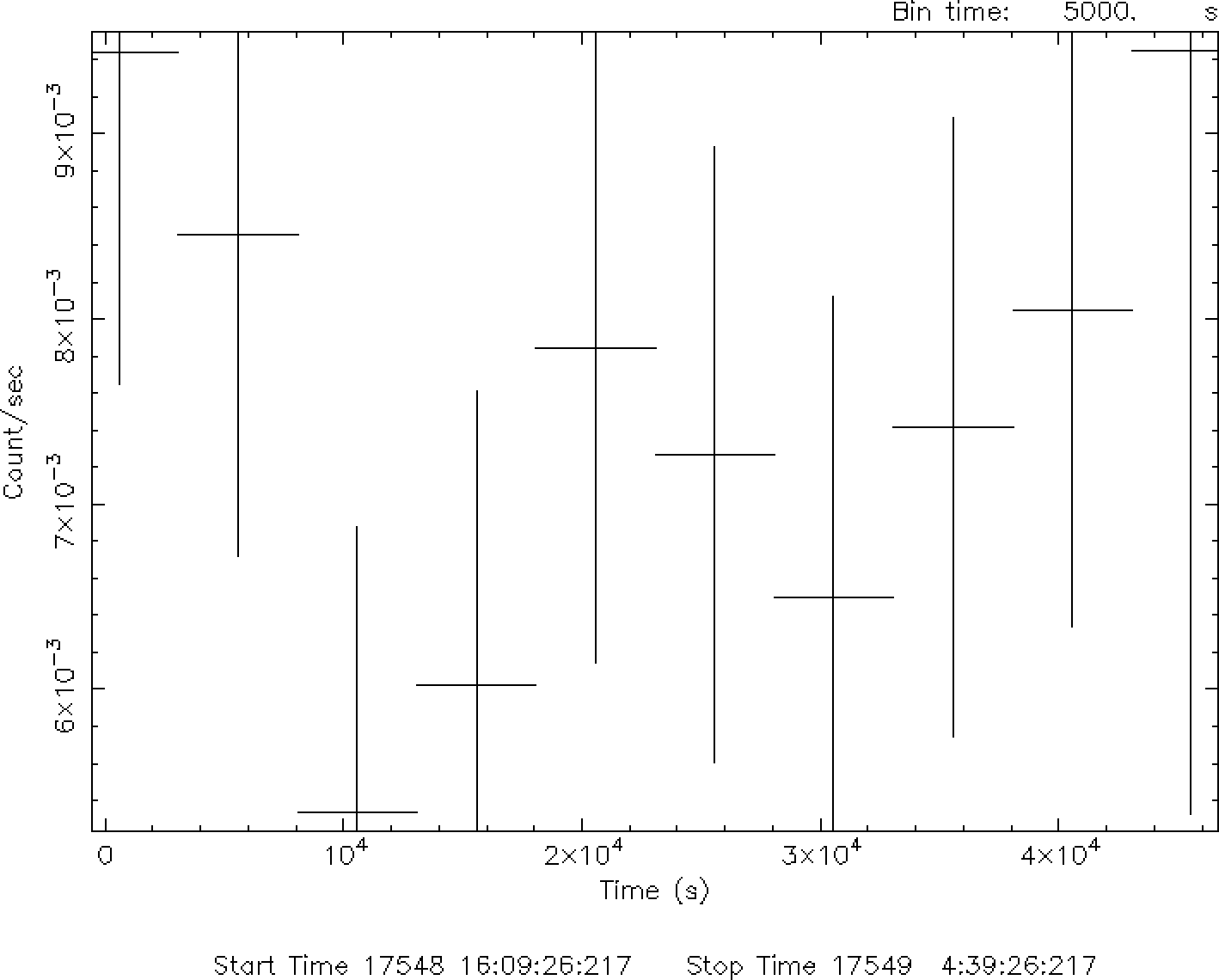}}
     \subfigure[X-05]{\label{X-05}\includegraphics[width=0.329\textwidth]{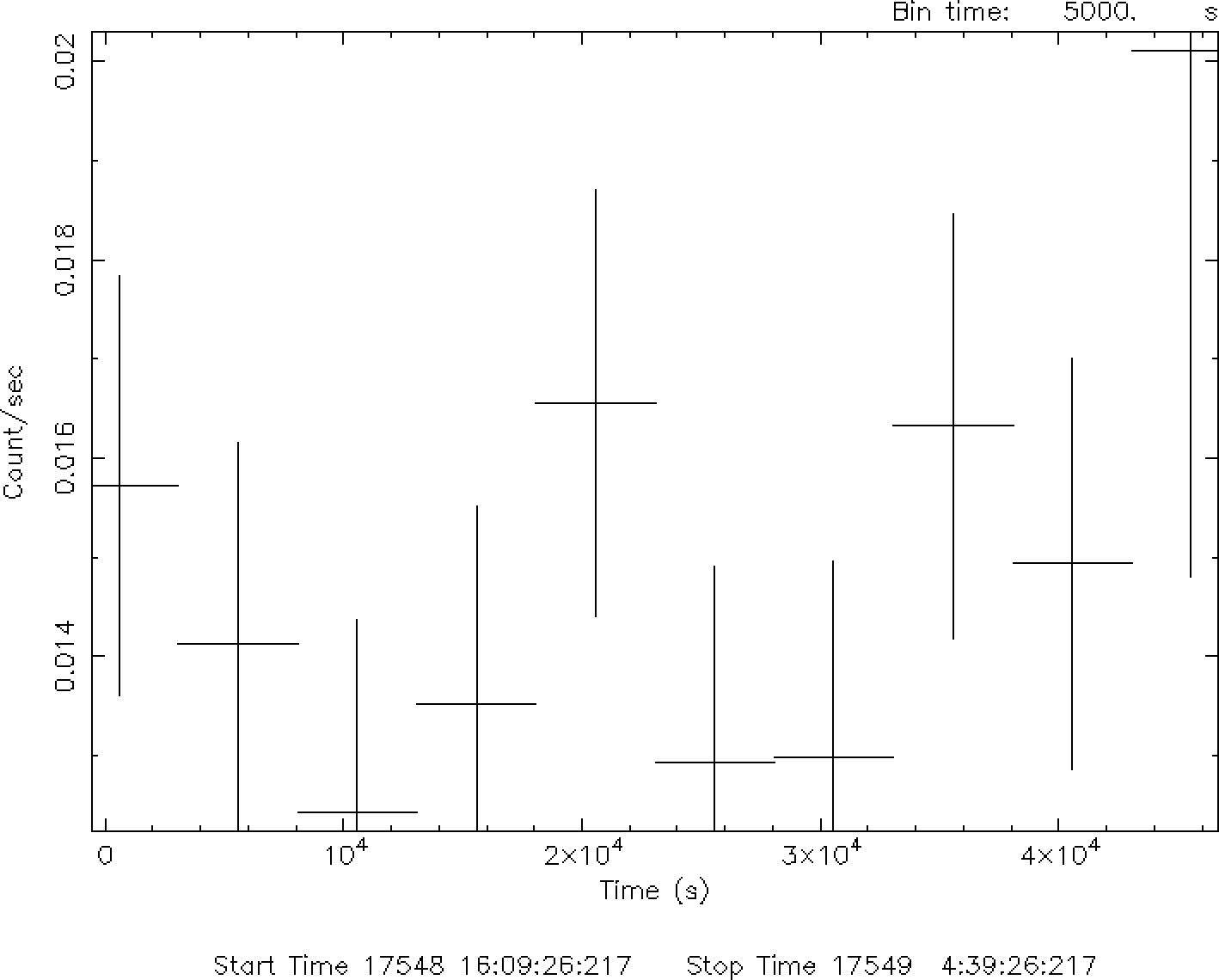}}
     \subfigure[X-07]{\label{X-07}\includegraphics[width=0.329\textwidth]{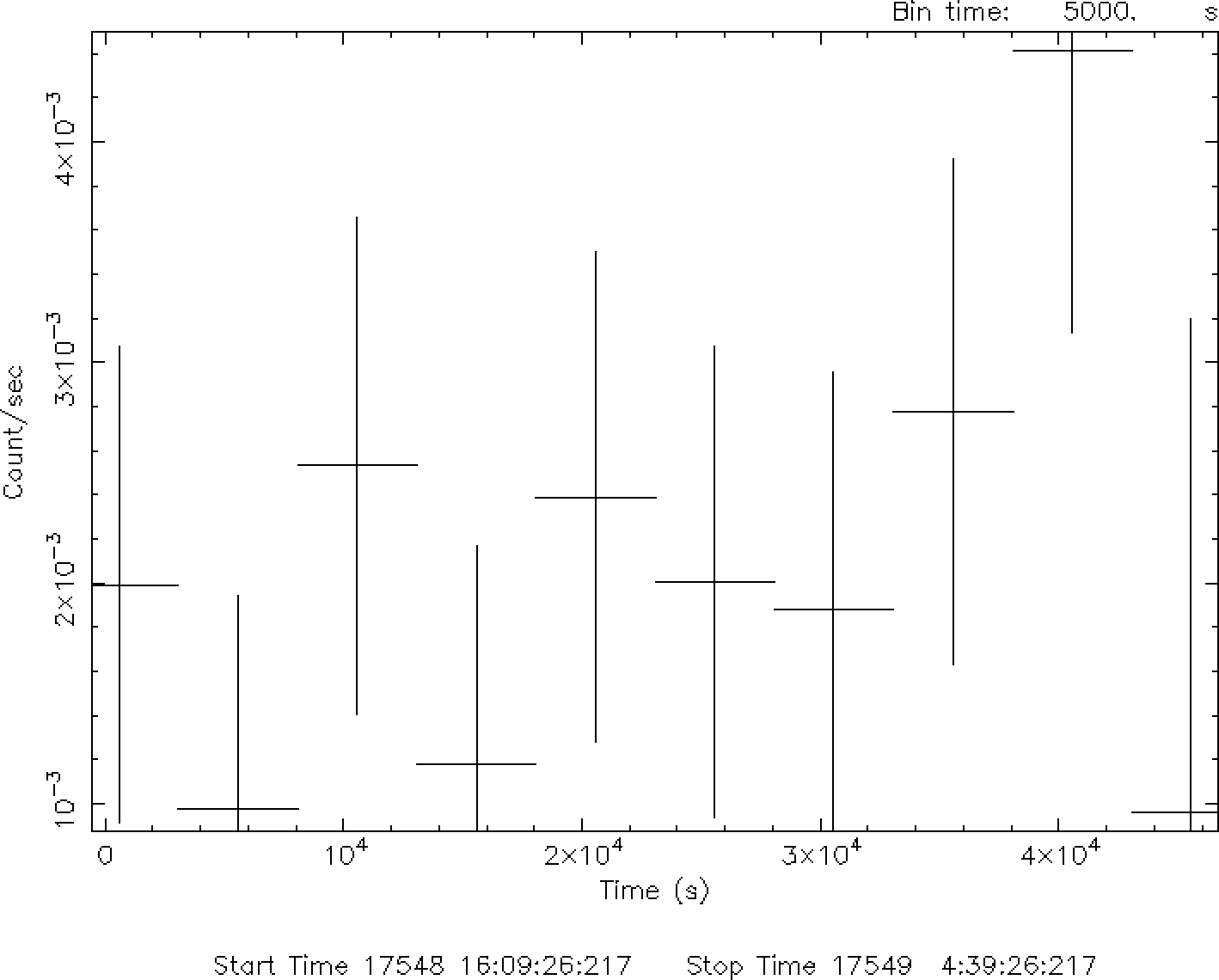}}
     \subfigure[X-08]{\label{X-08}\includegraphics[width=0.329\textwidth]{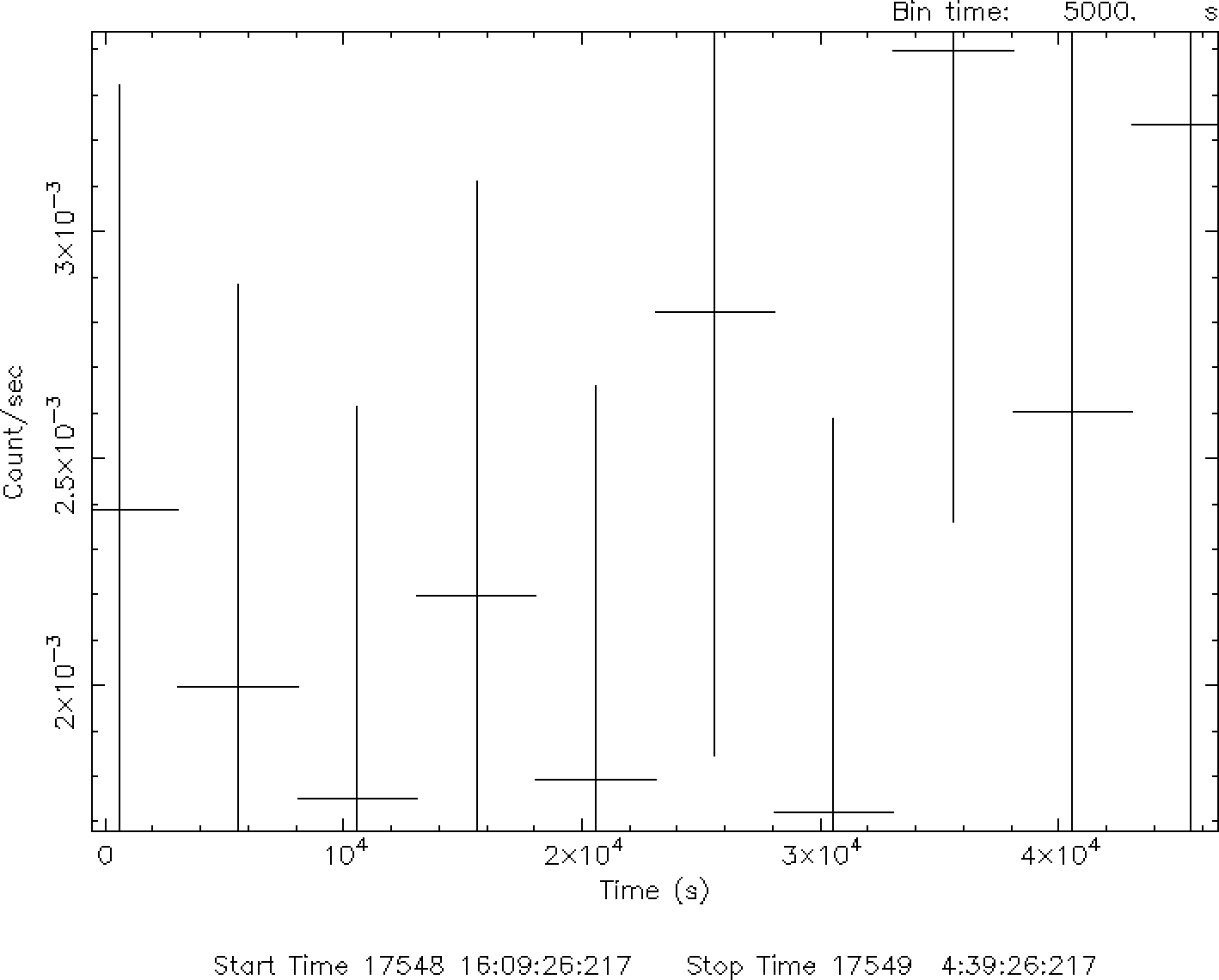}}
     \subfigure[X-14]{\label{X-14}\includegraphics[width=0.329\hsize]{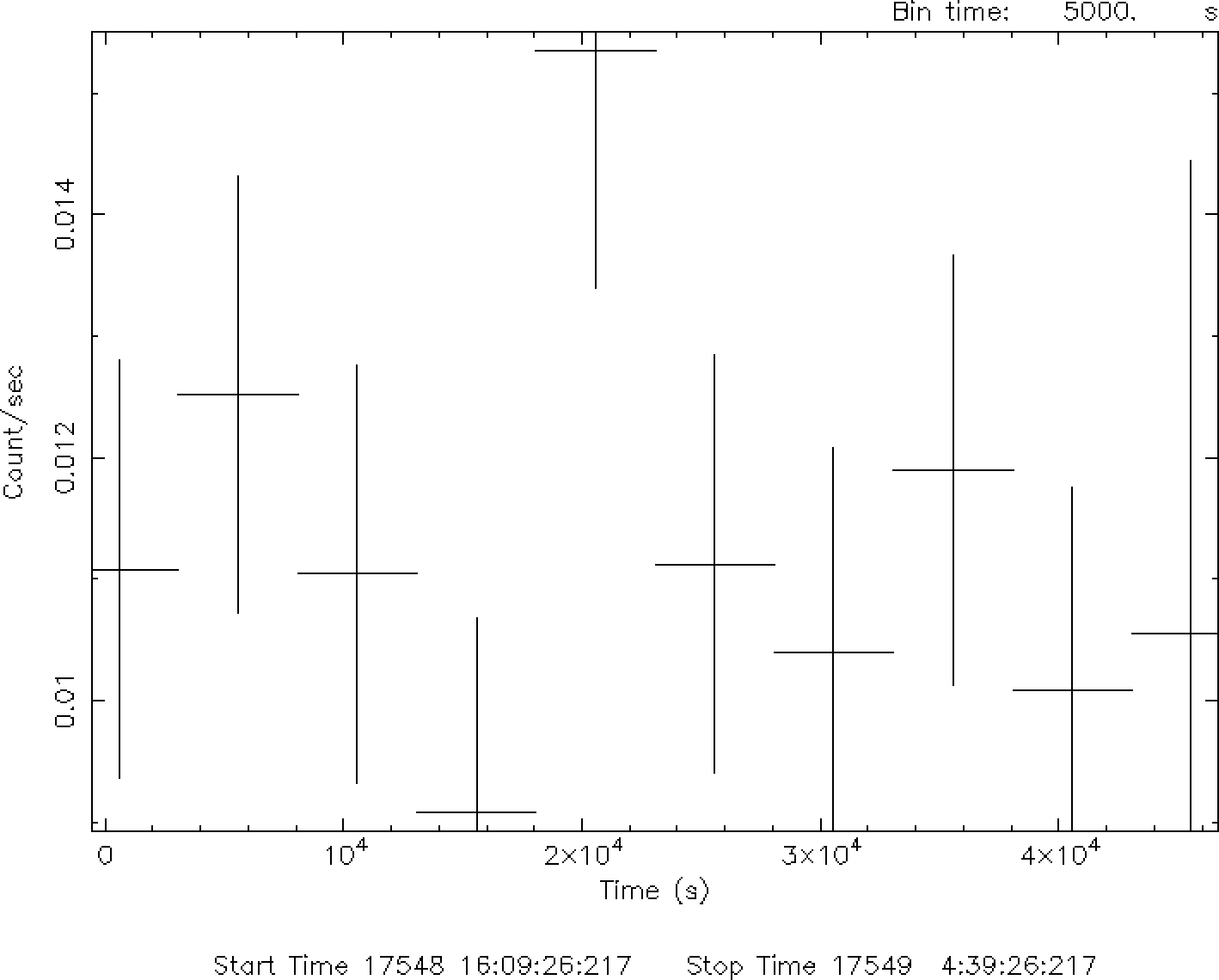}}
     \subfigure[X-17]{\label{X-17}\includegraphics[width=0.329\hsize]{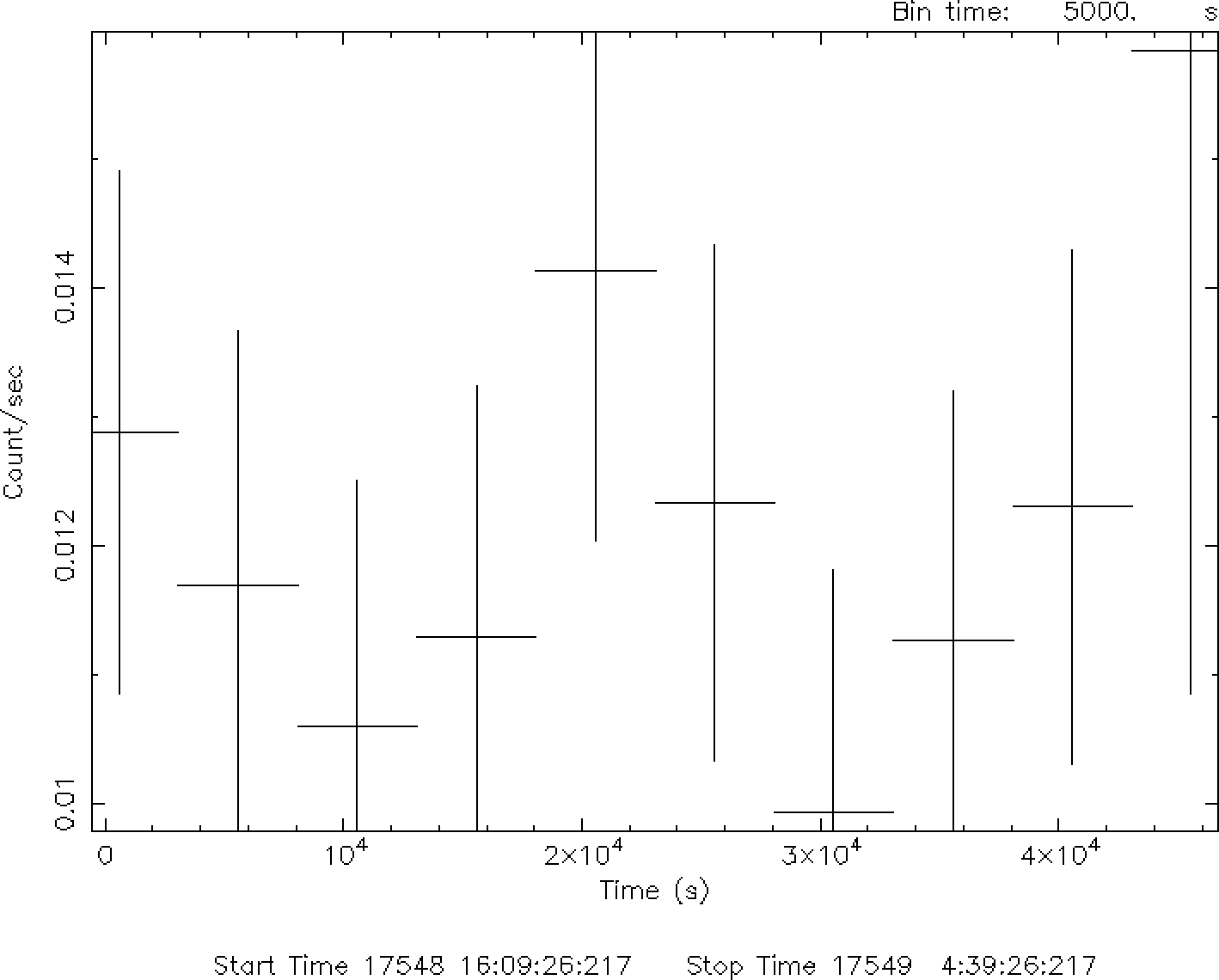}}
     \subfigure[X-18]{\label{X-18}\includegraphics[width=0.329\textwidth]{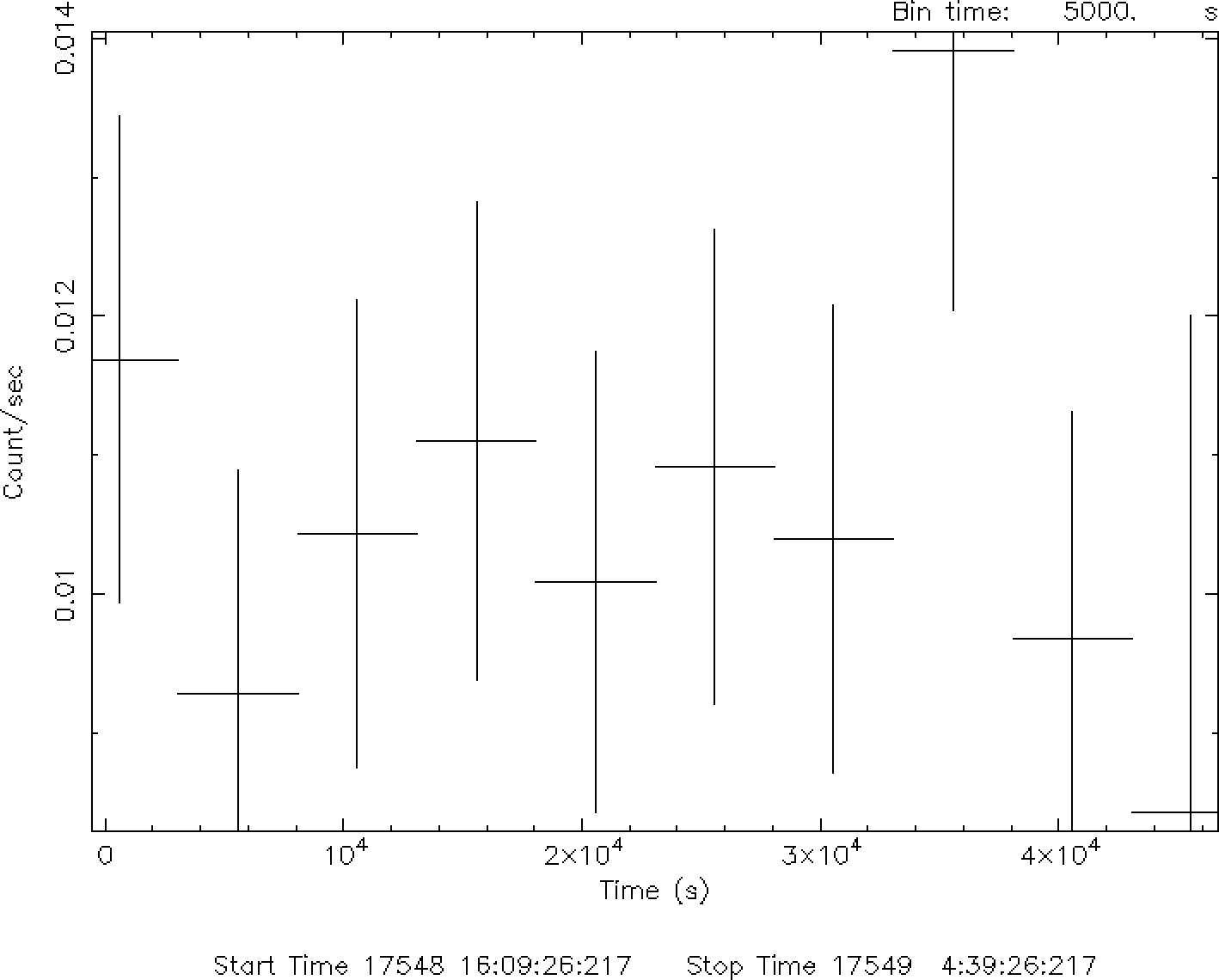}}
     \subfigure[X-24]{\label{X-24}\includegraphics[width=0.329\textwidth]{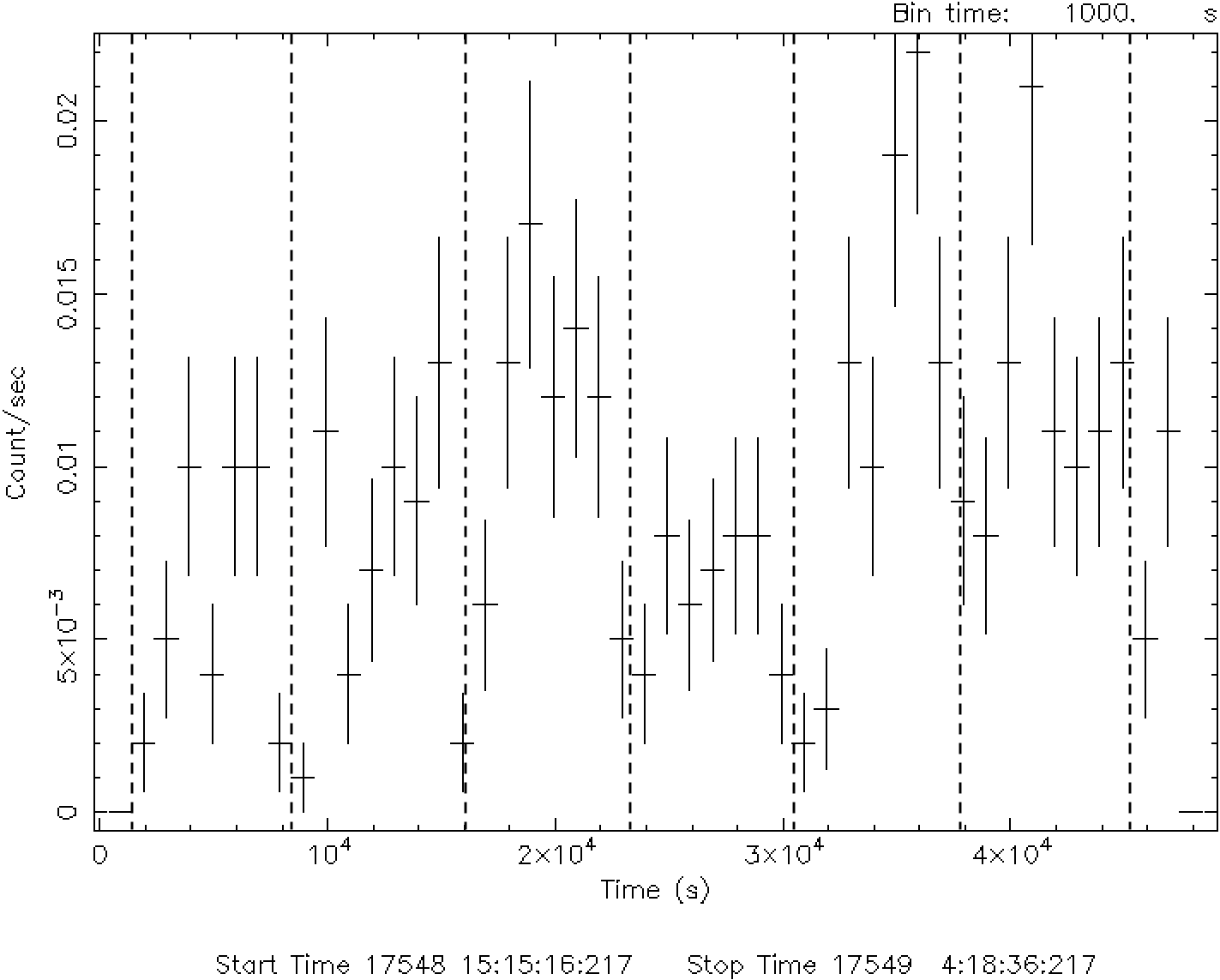}}
     \caption{These nine diagrams shows the light curves of the sources in Group A. 7 of them are ULXs which are X-01, X-03, X-05, X-14, X-17, X-18 and X-24. Their net count numbers are all larger than 300. The net count numbers of the two rest, X-07 and X-08, are both less than 300. The axes of abscissa are the exposure time from 0 to 45-ks. The axes of ordinate are count rates. The periodicities of these light curves are not obvious except for X-24. Due to its periodicity, we show it in Fig \ref{X-24} with auxiliary dotted line to indicate each cycle.}
     \label{lc-01}
     \end{center}
\end{figure*}

The light curves of the sources in Group A are presented in Fig \ref{lc-01}. 7 of them are ULXs whose net count number are all larger than 300. Most of them have no clear variability. One of the major reasons is the deficiency of the source counts, such as X-07 and X-08 (Fig \ref{X-07} and Fig \ref{X-08}). Among all, X-24 (Fig \ref{X-24}) has a relatively obvious variability. 

\begin{figure}[t]
      \begin{center}
      \includegraphics[width=\hsize]{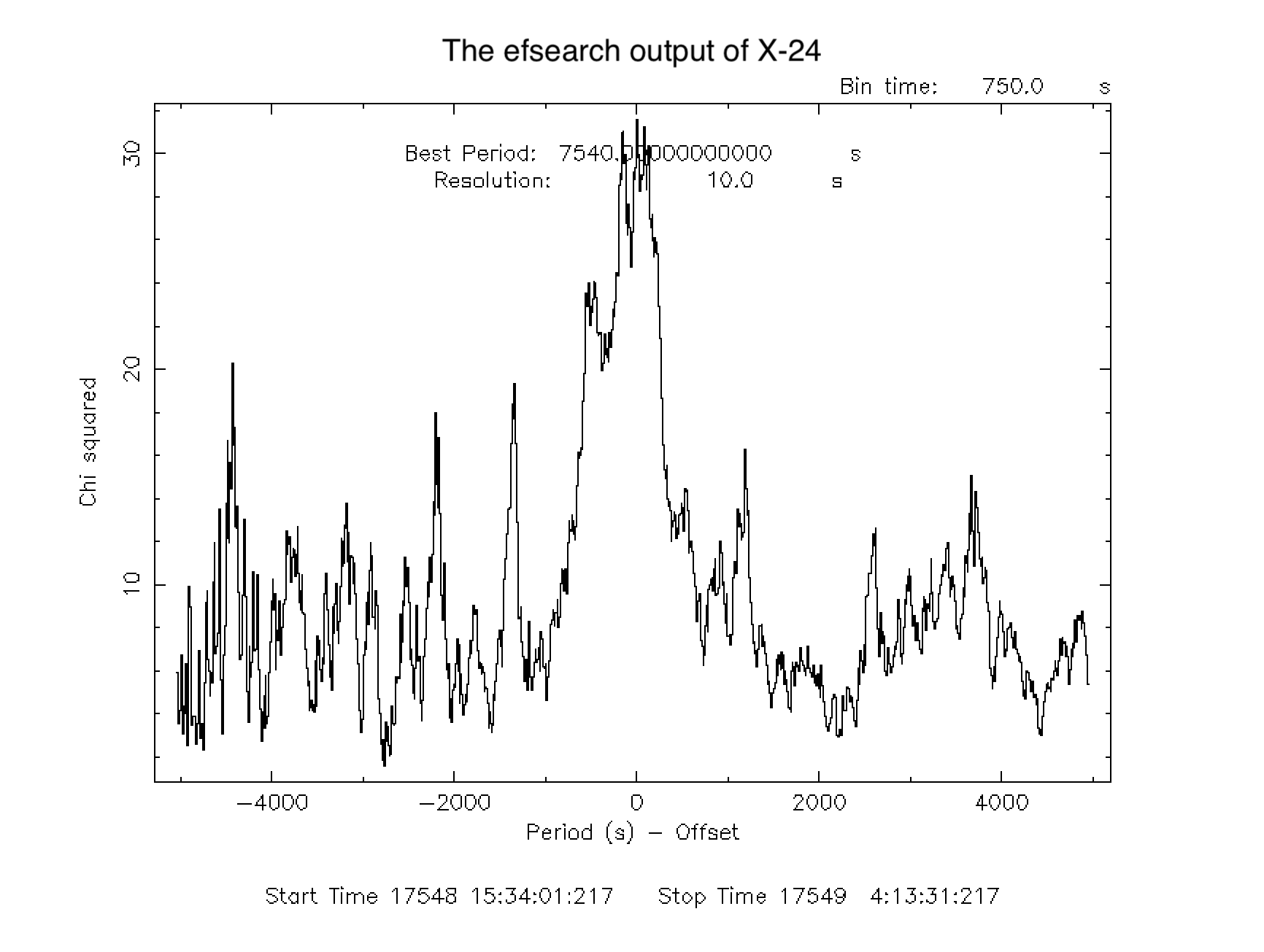}
      \caption{The period, $\approx$7500s, of X-24 searched by the task, \texttt{EFSEARCH}. It is the result for X-24 produced by the task, \texttt{EFSEARCH}. The axis of abscissa has been shifted to position the peak at zero point. Thus, the apparent peak in the middle point indicate the best possibility of period, $\approx$7500s.}
      \label{fig:3}
      \end{center}
   \end{figure}

\begin{figure}[t]
      \begin{center}
      \includegraphics[width=\hsize]{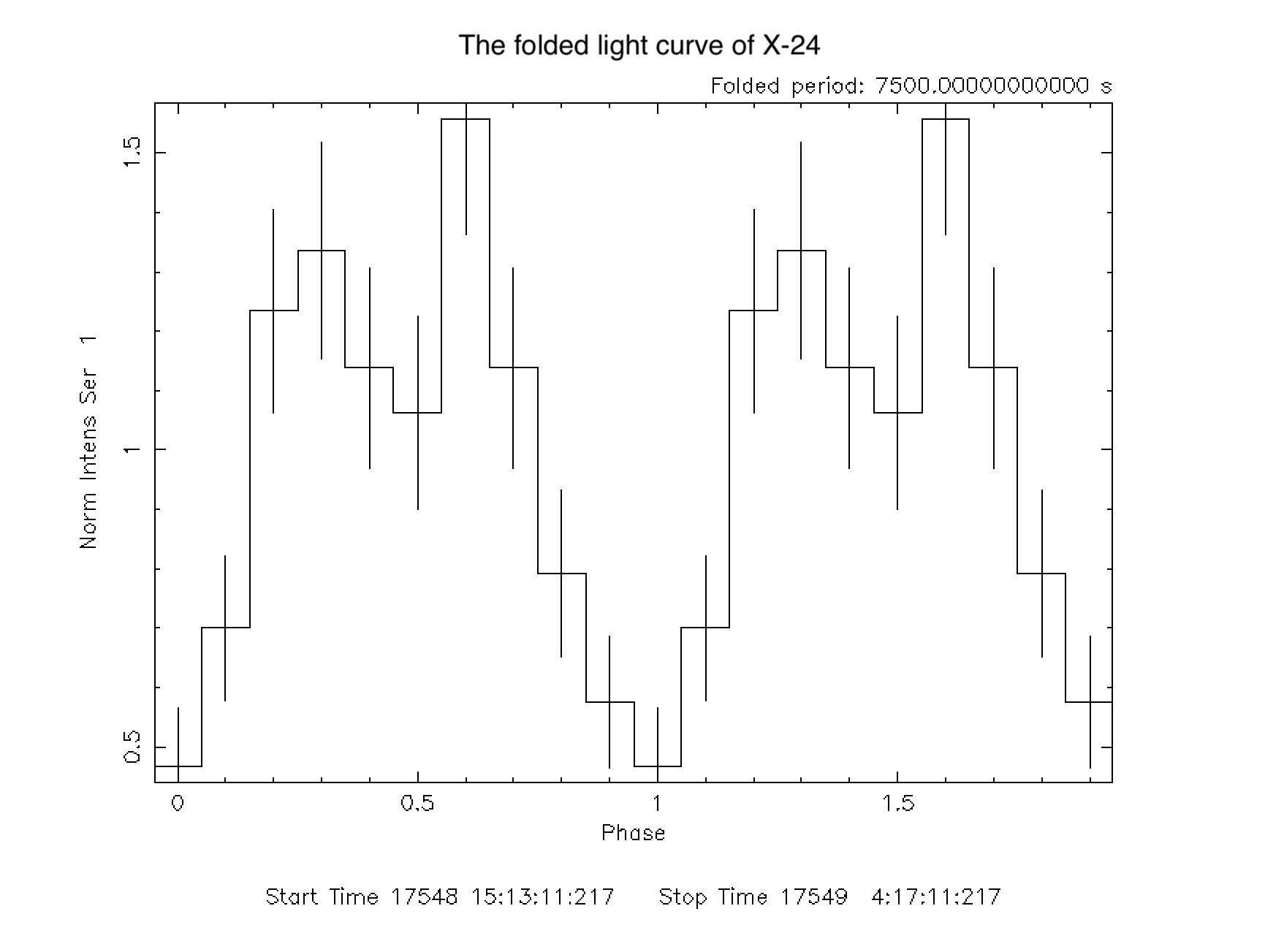}
      \caption{The figure shows the folded light curve of X-24 with 7500s, whose periodic pattern is relatively clearer than the un-folded light curve (Figure \ref{X-24}).}
      \label{fig:4}
      \end{center}
\end{figure}

For further confirmation of the periodicity, we perform timing analysis on all sources by the task, \texttt{EFSEARCH}. We search for the period from 1000s to 22000s using 200 resolution. The result of \texttt{EFSEARCH} is a periodogram which presents the $\chi ^2$ in different folding period. The \texttt{EFSEARCH} periodograms of all sources have no clear signals except X-24, whose periodogram presents three possible signals, which are $P\approx$ 21700s, 15600s and 7500s. All three signals have $\chi ^2$ greater than 30. The longest one (i.e., 21670s) is probably red noise given that it is almost half of the observing time of the \textit{Chandra} data. In contrast, the two shorter signals can be both real as the 15580-s periodicity is likely the alias of the 7500-s period. This can be demonstrated by the X-ray light curves phased on 7500s and 15600s, which have single peak and double peaks, respectively (Figure \ref{fig:3}). In this context, 7500s is the fundamental period of the ULX system. The $\chi^2$ of $P\approx7500$s is 35.21 with 9 degrees of freedom. This corresponds to a $p$-value of $1.8\times 10^{-4}$ for the $\chi^2$ distribution. We estimated the trails factor by considering the number of independent searches (defined by the Fourier resolution, $P^2/T$, where $P$ is the trail period and $T$ is the total exposure time of the data) in the search, which is around 47. Taking the number of trails into account, the chance probability is 0.8\% and the significance of the signal detection is approximately $2.7\sigma$ (two tailed). We examined the phased light curve of $P=7500$s and a quasi-sinusoidal profile is shown (Figure \ref{fig:4}). To check whether the profile was produced by short-term variabilities (e.g., flares on $\sim$1000-s timescale), the unfolded light curve was also investigated (Figure \ref{X-24}). X-ray dips that correspond to the phase zero of the phased light curve can be clearly seen over the entire exposure.
   
\begin{figure}[h!]
      \begin{center}
      \includegraphics[width=\hsize]{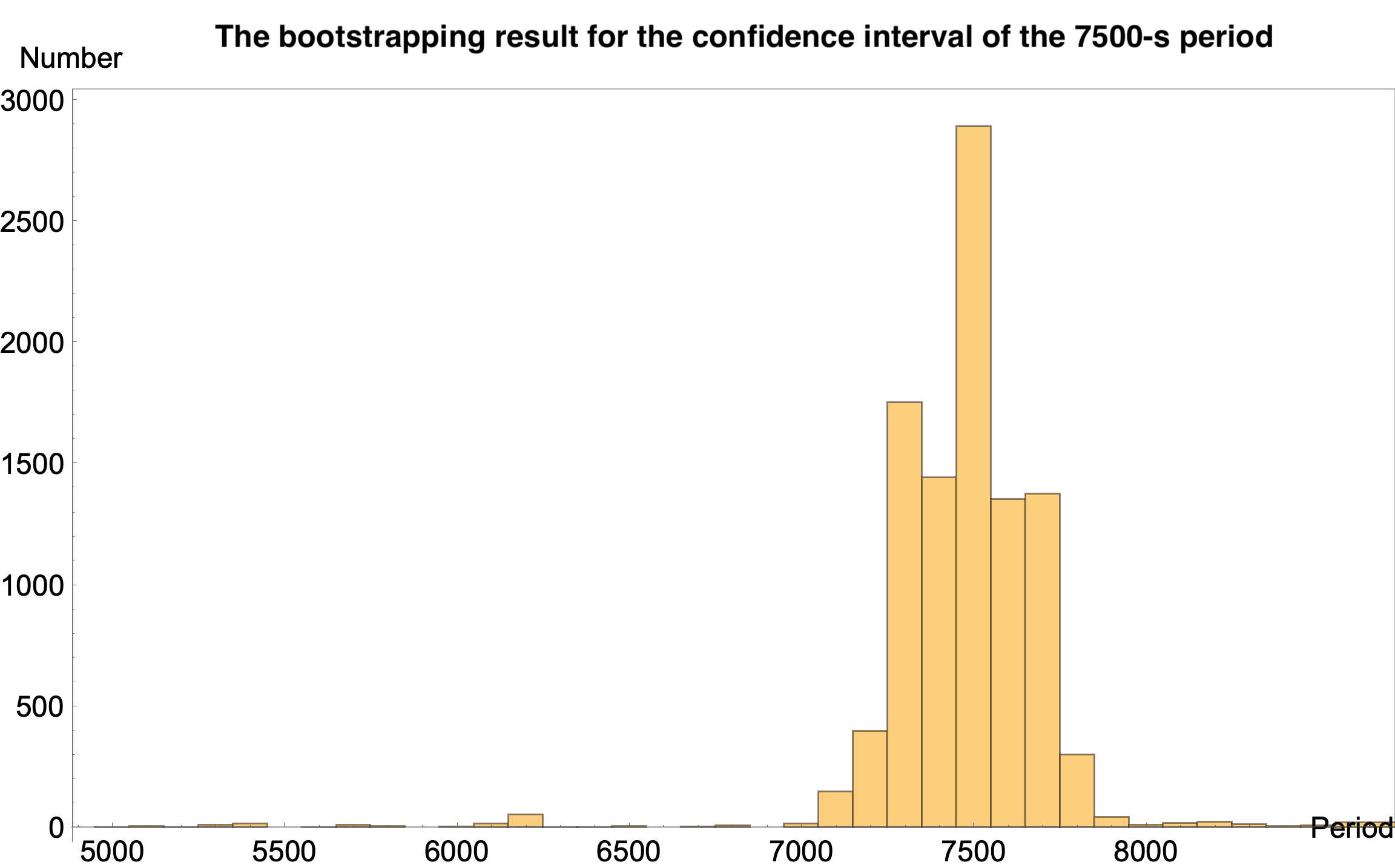}
      \caption{The distribution of the best-fit periods of the 10000 simulated X-ray light curves generated based on Poisson distribution. The median of the periods is 7500s and the upper and lower limit of 90\% confidence interval are 7700s and 7200s, respectively.}
      \label{fig:5}
      \end{center}
   \end{figure}

We estimated the confidence interval of the 7500-s period with bootstrapping. Based on the observed \textit{Chandra} light curve, 10000 sets of simulated data were produced assuming the Poisson distribution as the random noise model. For each simulated light curve, we ran \texttt{EFSEARCH} on it, and then picked up the peak in the range from 5000s to 9000s. Note that a smaller range of period was used here to avoid the two aforementioned signals above $P=10000$s and also save computing time. The median value of the peak distribution is 7500s with upper and lower limits of 90\% confidence interval at around 7700s and 7200s, respectively (Figure \ref{fig:5}).

   %----------------------------------------------------------------- 
\subsection{The observation of \textit{Swift} and XMM-Newton}

\begin{figure}[t]
     \begin{center}
	\includegraphics[width=0.95\hsize]{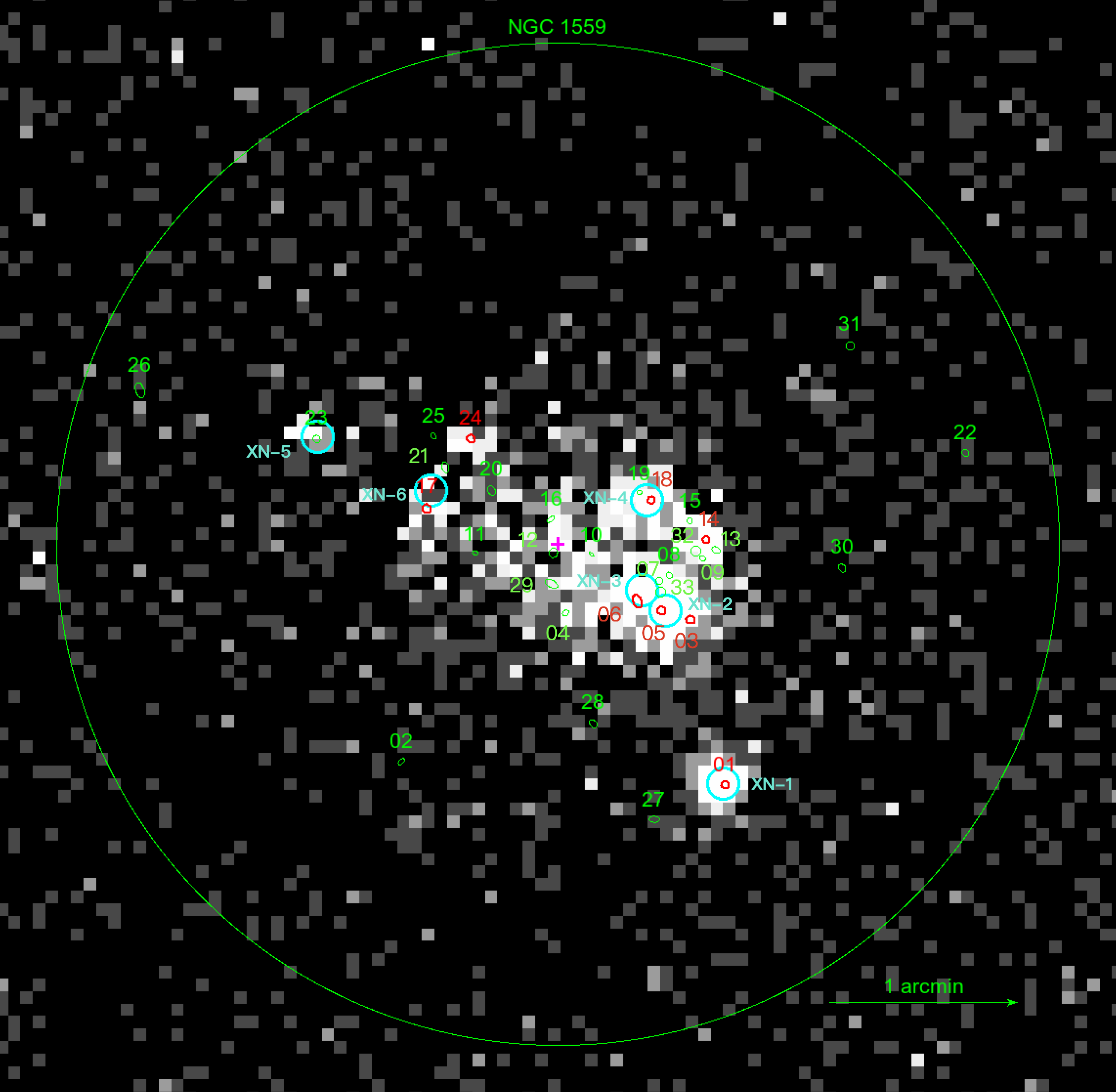}
     \caption{The image of the sources detected by \textit{XMM-Newton}. The image was taken by \textit{XMM-Newton}, whose center is indicated by the magenta cross-hairs at $\alpha\mathrm{(J2000)}=04^\mathrm{h}17^\mathrm{m}35\fs77$, $\delta\mathrm{(J2000)}=-62\arcdeg 47\arcmin 1\farcs2$. The cyan circles are the sources detected by \textit{XMM-Newton}. The green and red circles are the sources detected by \textit{Chandra}. Especially, the red circles mean the ULXs.}
     \label{xmm1}
     \end{center}
\end{figure}

\begin{figure}[t]
     \begin{center}
    \includegraphics[width=0.95\hsize]{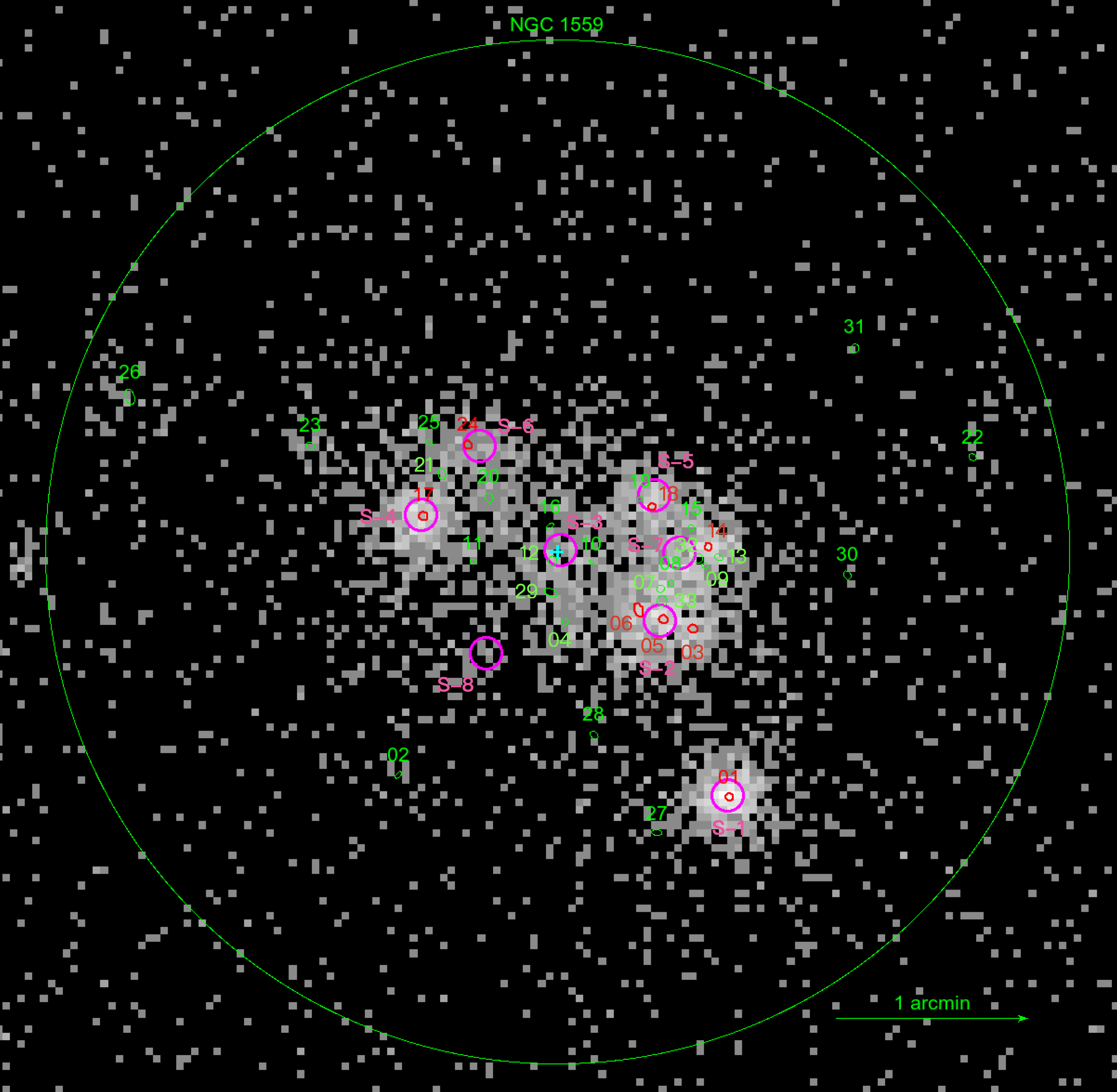}
     \caption{The image of the sources detected by \textit{Swift}. The image was taken by \textit{Swift}, whose center is indicated by the cyan cross-hairs at $\alpha\mathrm{(J2000)}=04^\mathrm{h}17^\mathrm{m}35\fs77$, $\delta\mathrm{(J2000)}=-62\arcdeg 47\arcmin 1\farcs2$. The magenta circles are the sources detected by \textit{Swift}. The green and red circles are the sources detected by \textit{Chandra}. Especially, the red circles mean the ULXs.}
     \label{Swift}
     \end{center}
\end{figure}

\textit{XMM-Newton} (14.4~ks; ObsID 0403070401) observed NGC~1559 in 2007 and \textit{Swift} observed from 2005 to 2016. According to the 4XMM-DR11 \citep{2020A&A...641A.136W} and 2SXPS \citep{2020ApJS..247...54E} catalogs, \textit{XMM-Newton} detected 6 sources in this observation and \textit{Swift} detected 8 sources in NGC~1559 up to now. Considering the lower resolutions of the observations, we assume that these \textit{XMM-Newton} and \textit{Swift} sources have an uncertainty radius of 5\arcsec. Figure \ref{xmm1} shows the X-ray image taken by \textit{XMM-Newton} where the 6 sources detected by \textit{XMM-Newton} and the 33 sources detected by \textit{Chandra} are both marked for comparison. 5 of the \textit{XMM-Newton} sources have possible counterparts in the \textit{Chandra} data and only one (XN-6) does not. As for \textit{Swift}, 6 out of the 8 sources are possibly related to some of the \textit{Chandra} sources (Figure \ref{Swift}). We summarize the properties of the cataloged XMM-Newton and the \textit{Swift}/XRT sources in Table \ref{tab:swift}.

%-------------------------------------------------------------------
    
\begin{figure*}[p]
     \begin{center}
    \includegraphics[width=0.96\hsize]{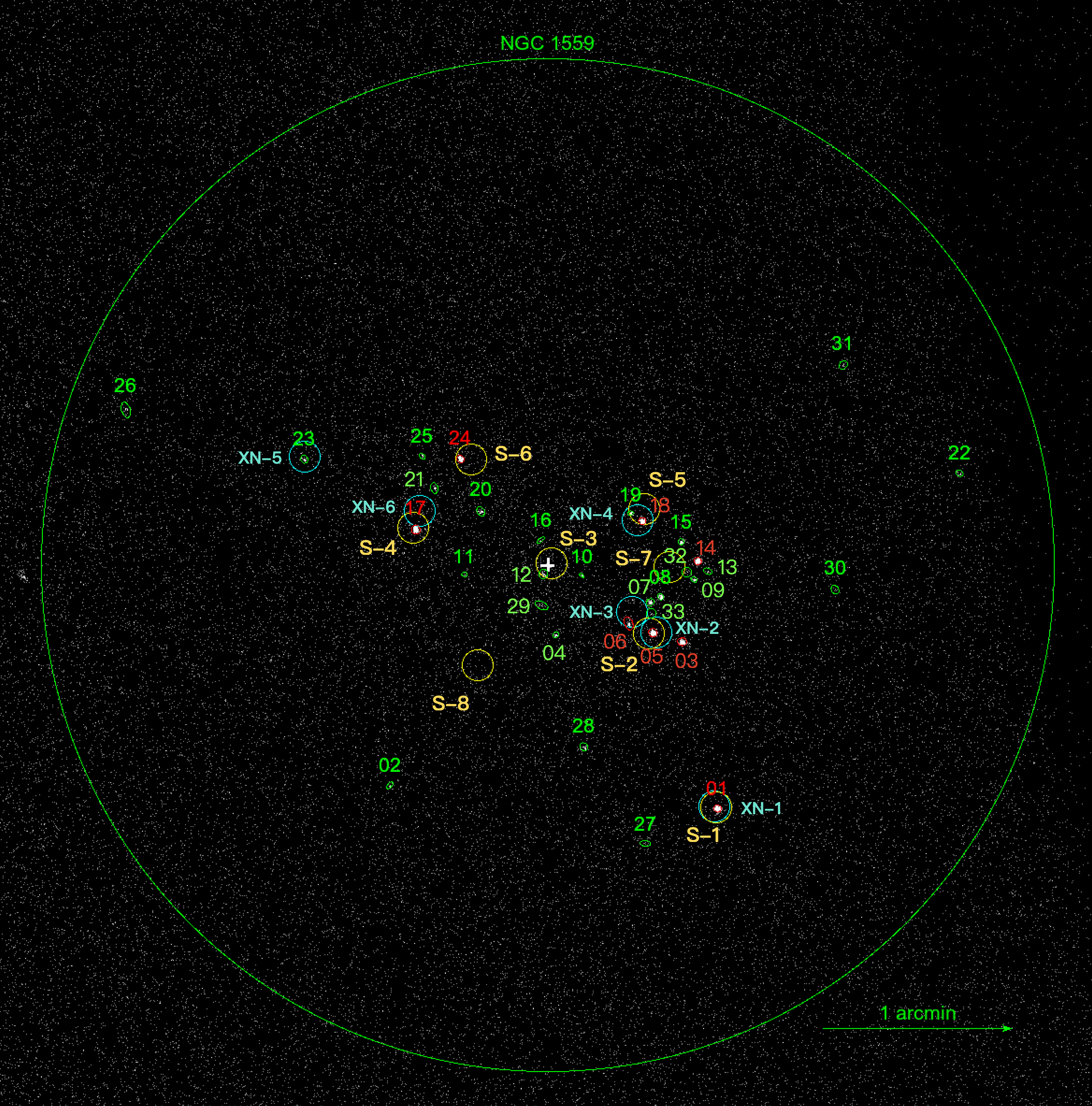}
     \caption{The image of the sources detected by \textit{Chandra}, \textit{XMM-Newton} and \textit{Swift}. The image was taken by \textit{Chandra}, whose center is indicated by the white cross-hairs at $\alpha\mathrm{(J2000)}=04^\mathrm{h}17^\mathrm{m}35\fs77$, $\delta\mathrm{(J2000)}=-62\arcdeg 47\arcmin 1\farcs2$. The cyan circles are the sources detected by \textit{XMM-Newton}. The yellow circles are the sources detected by \textit{Swift}. The green and red circles are the sources detected by \textit{Chandra}. Especially, the red circles mean the ULXs.}
     \label{all}
     \end{center}
\end{figure*}

In Figure \ref{all}, we place all the X-ray sources detected by \textit{Chandra}, \textit{XMM-Newton}, and \textit{Swift}/XRT on the \textit{Chandra}/ACIS image. We note that three \textit{XMM-Newton}/\textit{Swift} sources, named XN-6, S-7, and S-8, were not detected by \textit{Chandra}. However, we suggest that the non-detections are simply because of the relatively low resolutions of \textit{XMM-Newton} and \textit{Swift}/XRT. For example, S-7 is considered as a single source from the observation of \textit{Swift}. However, the high-resolution \textit{Chandra} observation can clearly show that it is composed of multiple neighboring sources. In conclusion, we find no evidence to show that these three X-ray sources were transients and were detected by \textit{XMM-Newton}/\textit{Swift}, but not \textit{Chandra}.

\subsubsection{\textit{Swift}/\textit{XMM-Newton} views on X-1/17/23/24}
As X-1, X-17, and X-23 are relatively bright and far away from the host galaxy (and hence less contaminated by the nearby sources and the diffuse emission of the host) and X-24 shows an interesting periodicity around 7500s, these 4 X-ray sources were selected for further investigations with the  \textit{Swift} and \textit{XMM-Newton} data. The observed properties of the X-ray sources are summarised in the following paragraphs, although X-17, X-23, and X-24 could still be affected by the contaminations, which we will also discuss below.

The \textit{Swift}/XRT observations (0.3--10~keV) has a total exposure time of 103~ks. All 4 X-ray sources can be seen in the X-ray image (Figure \ref{Swift}). Compared to the \textit{Chandra} spectra, the XRT-observed spectra of X-17, X-23 and X-24 are softer. We fit all the XRT spectra with the absorbed power-law model as we did for the \textit{Chandra} datasets, and the best-fit results are shown in Table \ref{tab:swift_xmm_spec}. The best-fit parameters of X-1 are generally consistent with the \textit{Chandra} result, the corrected flux is slightly higher though. However, all other three sources have much smaller intrinsic \nh\ values (the intrinsic \nh\ parameters are poorly constrained and only 95\% upper limits can be obtained from the XRT data). This corresponds to the aforementioned soft excesses, which compensate the absorbed emission and depress the best-fit \nh. We suspect that the soft excesses are not astrophysical, but very likely due to the contaminations from the nearby point sources and/or diffuse emission. As a result, we suggest that the XRT spectral results of X-17, X-23, and X-24 may not be reliable. We examined the XRT light curves\footnote{{Given the low count rates and the sparseness of the observations, the binning factors are 50~ks and 5000~ks for X-1/17 and X-23/24, respectively.}} for long-term variability (i.e., $\gtrsim1$~day). X-1 likely has mild X-ray variability on a timescale $\sim$10--100~days, but the rest of the sources do not show significant variability. For X-24, we also folded the XRT light curve on the $\approx7500$s, and a marginal variation can be seen (i.e., $\chi_\nu^2=11.4/7$). Given that the dataset is not clear as previously mentioned, we did not carry out further investigate on the possible variability.

\begin{figure}
     \begin{center}
    \includegraphics[angle=-90,width=0.45\textwidth]{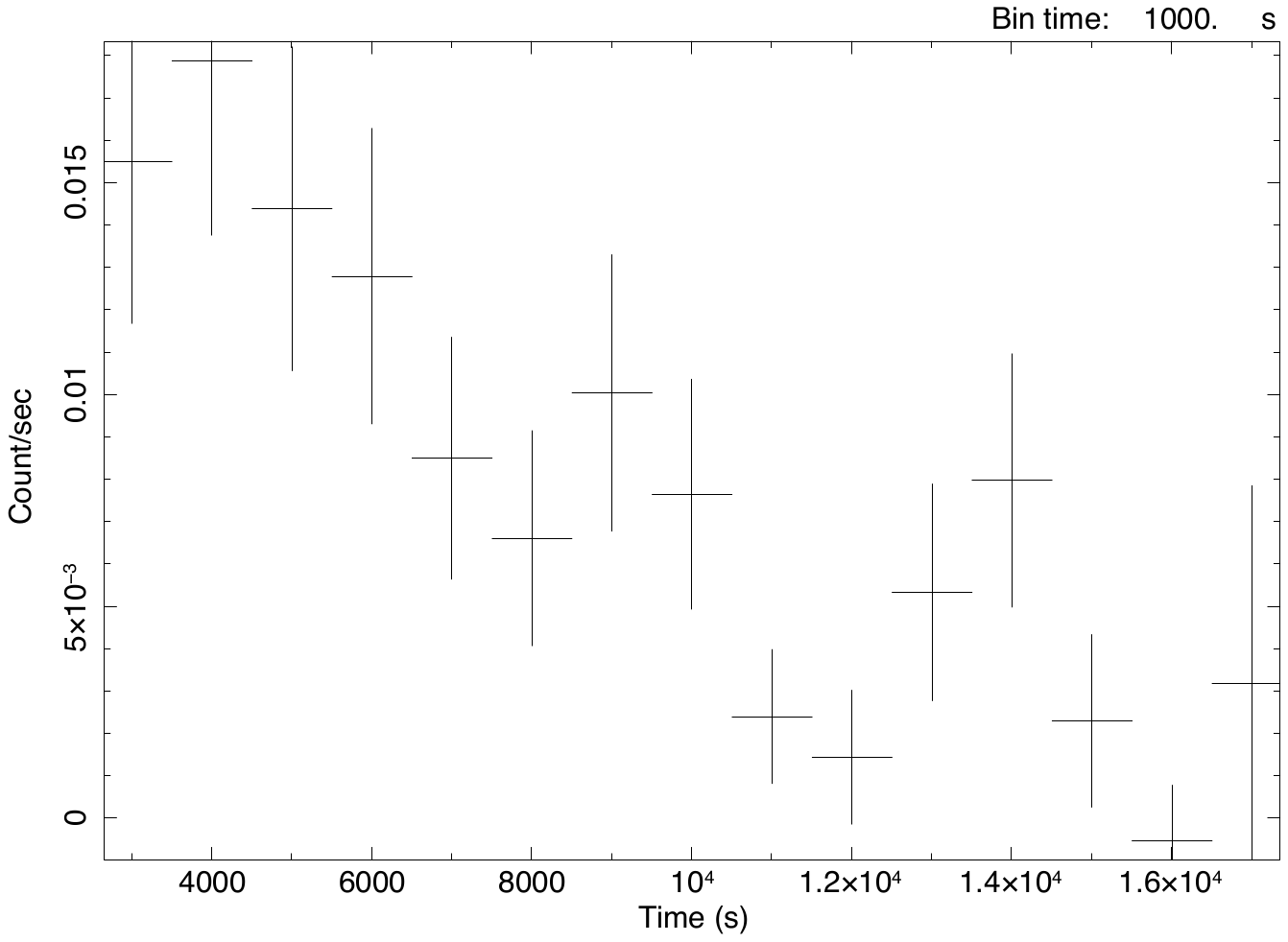}
     \caption{The \textit{XMM-Newton} light curve of X-24, which shows X-ray variability on a timescale of $\sim$5000--10000s (time zero: the start time of the \textit{XMM-Newton} observation). We caution that X-24 is likely contaminated by the nearby emission, and therefore, the variability may not be significant.}
     \label{fig:x24_xmm_lc}
     \end{center}
\end{figure}

For the \textit{XMM-Newton} observations (0.2--10~keV), the MOS-1/2 spectra of X-1 shows a hard photon index in the absorbed power-law model, but it is still consistent to the \textit{Chandra}/\textit{Swift} values considering the parameter uncertainties. The corrected flux of X-1 is fainter comparing to the \textit{Chandra}/\textit{Swift} observations, indicating that X-1 has long-term variability as suggested by the long-term XRT light curve. Similar to the XRT observations, X-17 and X-24 also show soft excesses and low \nh\ in the MOS-1/2 spectra, and therefore they are likely contaminated in the MOS-1/2 observations as well. In contrast, X-23 does not have the soft excess, and the best-fit intrinsic \nh\ is consistent with that observed by \textit{Chandra}, probably indicating a lower background contamination. However, given the very soft photon index of the MOS-1/2 dataset (i.e., $\Gamma=3.5^{+1.6}_{-1.2}$), whether the spectra are contaminated by the nearby soft diffuse emission is still questionable. For the light curves (binning factor of 1~ks), X-24 is the only source that shows possible varying X-rays, and no significant variability can be seen in the MOS-1/2 light curves of X-1, X-17, and X-23. The variability of X-24 is on a timescale of $\sim$5000--10000s (Figure \ref{fig:x24_xmm_lc}). This timescale is consistent with the 7500-s periodicity, and hence, they could be related. However, the variability could also be from the background noise as X-24 is likely contaminated by the nearby emission as mentioned.

From the \textit{XMM-Newton}/\textit{Swift} data analysis, we conclude that some X-ray sources were detected in the \textit{XMM-Newton}/\textit{Swift} observations. Except for X-1 (which is bright and also located at the outskirt of the host galaxy), the extracted spectral/temporal information of the sources is generally unusable because of the low spatial resolutions of the telescopes. Therefore, we will focus on the results of the \textit{Chandra} dataset in the following sections.

\begin{table*}
    \caption{Best-fit spectral parameters of the \textit{Swift}/XRT and \textit{XMM-Newton} MOS-1/2 spectra.}
    \label{tab:swift_xmm_spec}
    \begin{center}    
    \begin{tabular}{l c c c c l c}       % centered columns (4 columns)
        \hline\hline                                      % inserts double horizontal lines
        Instrument & ID & \nh\  & $\Gamma$ & $\log_{10}(\rm{Flux})$ & Statistic/dof & Note\\
         & & ($10^{22}$\,\cm) & & ($10^{-13}$\flux) & ($\chi^2$ or $C$) \\
        \hline
        XRT & X-1 & $0.21^{+0.11}_{-0.09}$ & $1.60^{+0.26}_{-0.24}$ & $-12.51^{+0.06}_{-0.05}$ & $\chi^2: 22.5/23$ & Long-term variability\\
        		& X-17 & $<0.05$ & $0.90^{+0.24}_{-0.21}$ & $-12.73^{+0.07}_{-0.08}$ & $\chi^2: 8.7/7$ & Likely contaminated\\
        		& X-23 & $<0.47$ & $2.14^{+1.6}_{-1.0}$ & $-13.57^{+0.63}_{-0.21}$ & $C: 27.9/27$ & Likely contaminated\\
        		& X-24 & $<0.17$ & $1.27^{+0.58}_{-0.43}$ & $-13.18^{+0.10}_{-0.11}$ & $C: 53.5/71$ & Likely contaminated\\
		\hline
        MOS-1/2 & X-1 & $<0.26$ & $1.08^{+0.65}_{-0.42}$ & $-12.73^{+0.09}_{-0.09}$ & $\chi^2: 17.9/7$\\
        		& X-17 & $<0.18$ & $1.31^{+0.47}_{-0.34}$ & $-13.07^{+0.09}_{-0.09}$ & $C: 133.3/110$ & Likely contaminated\\
        		& X-23 & $0.56^{+0.41}_{-0.30}$ & $3.5^{+1.6}_{-1.2}$ & $-12.59^{+1.00}_{-0.52}$ & $C: 60.1/63$ & Could be contaminated\\
        		& X-24 & $<0.13$ & $1.26^{+0.44}_{-0.25}$ & $-13.05^{+0.08}_{-0.09}$ & $C: 115.9/122$ & Likely contaminated\\
    		\hline
    \end{tabular}
    \end{center}
\tablecomments{The spectral model is the same absorbed power-law model used for the \textit{Chandra} data (Table \ref{tab:1}). The X-ray fluxes (0.3--7.0~keV) have been corrected for absorption. We comment that X-17, X-23, and X-24 could be contaminated by the nearby point sources and/or diffuse emission because of the insufficient spatial resolutions.}
    \end{table*}

\section{Discussions} \label{sec:discussions}
\subsection{NGC~1559 as a ULX factory}
In NGC~1559, there are 33 X-ray sources detected, including 8 ULXs. It is possible that some of these 8 ULXs are in fact distant sources, like active galactic nuclei (AGNs). Therefore, we made use of the Chandra Deep Field-South survey \citep{2017ApJS..228....2L} to estimate the number of the background sources in our ULX list. Using the X-ray flux of the faintest ULX in the list as the reference level (i.e., $10^{-13.25}$ \flux), we found the expected number of background source is $\sim1$. Thus, the number of bonafide ULXs is around 7. \cite{2020MNRAS.498.4790K} proposed a function to quantify the number of ULXs in late type galaxy (LTGs):
\begin{equation}
N_{ulx}=0.45^{+0.06}_{-0.09}\times\frac{\text{SFR}}{\text{M}_\sun\text{yr}^{-1}}+3.3^{+3.8}_{-3.2}\times\frac{\text{TSM}}{\text{10}^{12}\text{M}_\sun}
\end{equation}
, where the SFR and TSM are the star forming rate and total stellar mass, respectively. In NGC~1559, the TSM is $10^{10.23}\,\text{M}_\sun$ reported in the \textit{Heraklion Extragalactic Catalogue} (\textit{HECATE}) and the SFR is $3.9\,\text{M}_\sun\cdot\text{yr}^{-1}$ from \cite{2013MNRAS.428.1927C}. Adopting these parameters, the $N_{ulx}$ is estimated to be $1.8^{+0.3}_{-0.4}$, much smaller than the number of bonafide ULXs $(\sim7)$. 
\begin{table*} [!t]
    \caption{The list of X-ray sources in 4XMM-DR11 and 2SXPS catalogs cross-checking with \textit{Chandra} data}
    \label{tab:swift}
    \renewcommand\arraystretch{1.2}
    \begin{center}    
    \begin{tabular}{c c c c c}       % centered columns (4 columns)
        \hline\hline                                      % inserts double horizontal lines
        \multirow{2}{*}{IAUName\,(Label)} & \multirow{2}{*}{R.A.} & \multirow{2}{*}{Dec.} & Flux                     & ID \\
                                                               &                                  &                                   & ($10^{-13}$\flux) & (Table \ref{tab:1}) \\
        \hline                                               % inserts single horizontal line
J041728.1-624817\,(XN-1) & 4:17:28.10 & -62:48:17.5 & 1.69 ($\pm$0.18) & 1 \\
J041730.7-624722\,(XN-2) & 4:17:30.74 & -62:47:22.4 & 3.02 ($\pm$0.24) & 5 \\
J041731.8-624715\,(XN-3) & 4:17:31.88 & -62:47:16.0 & 10.17 ($\pm$0.63) & 6 \\
J041731.6-624647\,(XN-4) & 4:17:31.64 & -62:46:47.2 & 0.82 ($\pm$0.12) & 18, 19 \\
J041746.9-624627\,(XN-5) & 4:17:46.93 & -62:46:27.0 & 0.23 ($\pm$0.05) & 23 \\
J041741.6-624644\,(XN-6) & 4:17:41.66 & -62:46:44.1 & 3.73 ($\pm$0.39) & $\times$ \\
J041728.0-624817\,(S-1) & 4:17:29.03 & -62:48:17.6 & 3.87 ($\pm$0.21) & 1 \\
J041731.1-624722\,(S-2) & 4:17:31.11 & -62:47:22.7 & 5.08 ($\pm$0.22) & 5 \\
J041735.6-624700\,(S-3) & 4:17:35.61 & -62:47:00.7 & 3.86 ($\pm$0.26) & 12 \\
J041741.9-624649\,(S-4) & 4:17:41.96 & -62:46:49.5 & 3.77 ($\pm$0.20) & 17 \\
J041731.3-624643\,(S-5) & 4:17:31.35 & -62:46:43.6 & 3.01 ($\pm$0.18) & 18,19 \\
J041739.2-624627\,(S-6) & 4:17:39.29 & -62:46:27.9 & 1.77 ($\pm$0.15) & 24 \\
J041730.1-624701\,(S-7) & 4:17:30.19 & -62:47:01.7 & 6.38 ($\pm$0.27) & $\times$ \\
J041739.0-624732\,(S-8) & 4:17:39.01 & -62:47:32.7 & 1.02 ($\pm$0.18) & $\times$ \\
    \hline                                   %inserts single line %04^\mathrm{h}17^\mathrm{m}35\fs77$, $\delta\mathrm{(J2000)}=-62\arcdeg 47\arcmin 1\farcs2
    \end{tabular}
    \end{center}
\tablecomments{This table lists the sources detected by \textit{XMM-Newton} and \textit{Swift} observations of NGC~1559 and we match these sources with the ones detected by \textit{Chandra}. Most of the sources have counterparts in  the \textit{Chandra} data and the \textit{Chandra} sources are listed in the last column. There are three sources with no \textit{Chandra} counterparts, one of which was observed by \textit{XMM-Newton} and the other two were observed by \textit{Swift}, possibly due to the low resolutions of the instruments.}
    \end{table*}
The excess of ULXs has been suggested to be associated with the low-metallicity of the galaxy \citep{2020MNRAS.498.4790K}. \cite{10.1111/j.1365-2966.2012.20595.x} gave the following polynomial fit with the optimum projection for metallicity:
\begin{equation}
Z=43.448-12.193x+1.373x^2-0.050x^3
\end{equation}
, where $Z=12+\log(\text{O}/\text{H})$ and $x=\log(\text{TSM})-0.19\log(\text{SFR})$. \cite{2020MNRAS.498.4790K} suggest that the lowest metallicities correspond to $Z=8.3-8.6$, the intermediate metallicities correspond to $Z=8.6-8.7$ and the near-solar metallicities $Z$ are larger than 8.7 ($Z_\sun=8.9$). Based on their statistics, The galaxies with the lowest metallicities may host more ULXs than that we expected. The galaxies of intermediate metallicity may have no excess of ULXs. If the metallicities are near the solar metallicity $Z_\sun$, the galaxies may host half of the expected ULXs. In the case of NGC~1559 $Z$ is almost 8.84, which is considered as the near-solar metallicity. Based on the deduction of \cite{2020MNRAS.498.4790K}, the amount of ULXs in NGC~1559 should just be half of the expected ULXs ($N_{ulx}\simeq1$). Nevertheless, it indicates an excess of ULXs. Thus, the speculation of low-metallicity for the excess of ULXs might not be appropriate for this galaxy.

Besides the low-metallicity, we have also considered the possibility that NGC 1559 was a starburst galaxy in the past. Like the case of M82 and M81, a pair of interacting galaxies, close encounters with neighboring galaxies could trigger star formation and significantly raise the SFR. However, the two closest galaxies to NGC~1559 are ESO084-015 and ESO084-013, and their distances to NGC~1559 are 3.21~Mpc and 5.91~Mpc, respectively. These distances are not as close as M82 and M81 (0.05 Mpc). 

In addition to the excess of ULXs, another feature of NGC 1559 is that it hosted many supernovae. Under the conditions of the similar TSM and SFR, NGC~6946 is a galaxy comparable to NGC~1559 with TSM and SFR of $10^{10.6}\,\text{M}_\sun$ and $2.6\,\text{M}_\sun\cdot\text{yr}^{-1}$, respectively, recorded in the \textit{HECATE}. Not only these values are fairly close to those of NGC~1559, NGC 6949 also hosts 10 supernovae. However, NGC~6946 only has 4 ULXs \citep{Earnshaw_2019}. Their characteristics are still slightly different.

%--------------------------------------------------------------------

\subsection{X-24}

In the \textit{Chandra} X-ray light curves of the 8 ULXs, only X-24 showed a possible periodic signal. Despite the fact that periodicities are not uncommon for ULXs, common timescales of ULXs are in the orders of seconds or days. For instance, ULX pulsars can spin on timescales of (sub-)seconds \citep{2014Natur.514..202B,2020A&A...642A.174M}, while daily periodicities can be the orbital periods of the binaries that host ULXs \citep{Kaaret_2007,2013Natur.503..500L}. In contrast, hourly periodicities like the one seen in X-24 are relatively rare. \cite{2002ApJ...581L..93L} reported an ULXs with a 2-hour period in M51 and suggested the ULXs might be a low mass X-ray binary with a dwarf companion and a compact accretor \citep{2002ApJ...581L..93L}. The only other known case of ULX varying on timescales of thousand seconds is M81 X-6, which is viewed as a pulsar. M81 X-6 has a period of 2681s, but its statistical significance  is just above 95\% \citep{10.1093/mnras/staa976}. We have investigated the latest reviews for ULXs, which did not mention this kind of 2-hour period \citep{2017ARA&A..55..303K,2004cosp...35.1354M,2004RMxAC..20...46F,2021AstBu..76....6F}.
%there are few other reports on the ULXs with 2-hour period, in our knowledge.

The 7500-s period is too long for a spin period of a pulsar, but too short for the super-orbital period of a ULX \citep{2020MNRAS.495L.139T}. Although the periodicity is much shorter than the orbital periods of many ULXs (e.g., 8.2 days for M101~ULX-1; \citealt{2013Natur.503..500L}), we suspect that X-24 is a compact X-ray binary and the 7500-s periodicity is the orbital period of the system. On this ground, we deduced that X-24 is a low mass X-ray binary (LMXB) and the mass of the companion star is $M_1\sim1{M}_\sun$.
In the following considerations, we therefore simplify the case by assuming $M_1=1{M}_\sun$. It is important to emphasize that we currently know very little regarding the mass of the companion, but the models with $M_1=1{M}_\sun$ could still give us clues to the physical origin of the system. The calculations with $M_1=1{M}_\sun$ would also be transformed to other mass scales more easily, if the companion mass could be accurately measured in the future.

Given that X-24 is a ULX and the accretion rate must be high, we considered a Roche-lobe overflow (RLOF) situation, in which the size of the donor exceeds the Roche-lobe resulting in a high rate of mass transfer. The size of Roche lobe ($r_{\rm rlof}$) depends on the mass ratio, $q=M_1/M_2=1/M_2$, of the binary system, where $M_2$ is the mass of accretor, and the relations can be written as \cite{1983ApJ...268..368E}.%\citep{1983ApJ...268..368E}. 

\begin{equation}
\frac{r_{\rm rlof}}{a}=\frac{0.49q^{\frac{2}{3}}}{0.6q^{\frac{2}{3}}+\ln(1+q^{\frac{1}{3}})},
      \label{eq:roche}
\end{equation}
where  $a=\left[ \frac{G(M_1+M_2)}{4\pi^2}\cdot P^2_{orb}\right]^{1/3}$.

Under the conditions of $P_{\rm orb}=7500$s and $M_1=1{M}_\sun$, we found that the radius of the Roche lobe is always smaller than $0.4R_\sun$, even a massive compact object (i.e., $100M_\sun$) was assumed. This means that RLOF is almost always possible, no matter the compact object is a neutron star, a stellar-mass BH, or an IMBH, as the low-mass companion is likely larger than $0.4R_\sun$

\begin{figure}[h]
      \begin{center}
      \includegraphics[width=\hsize]{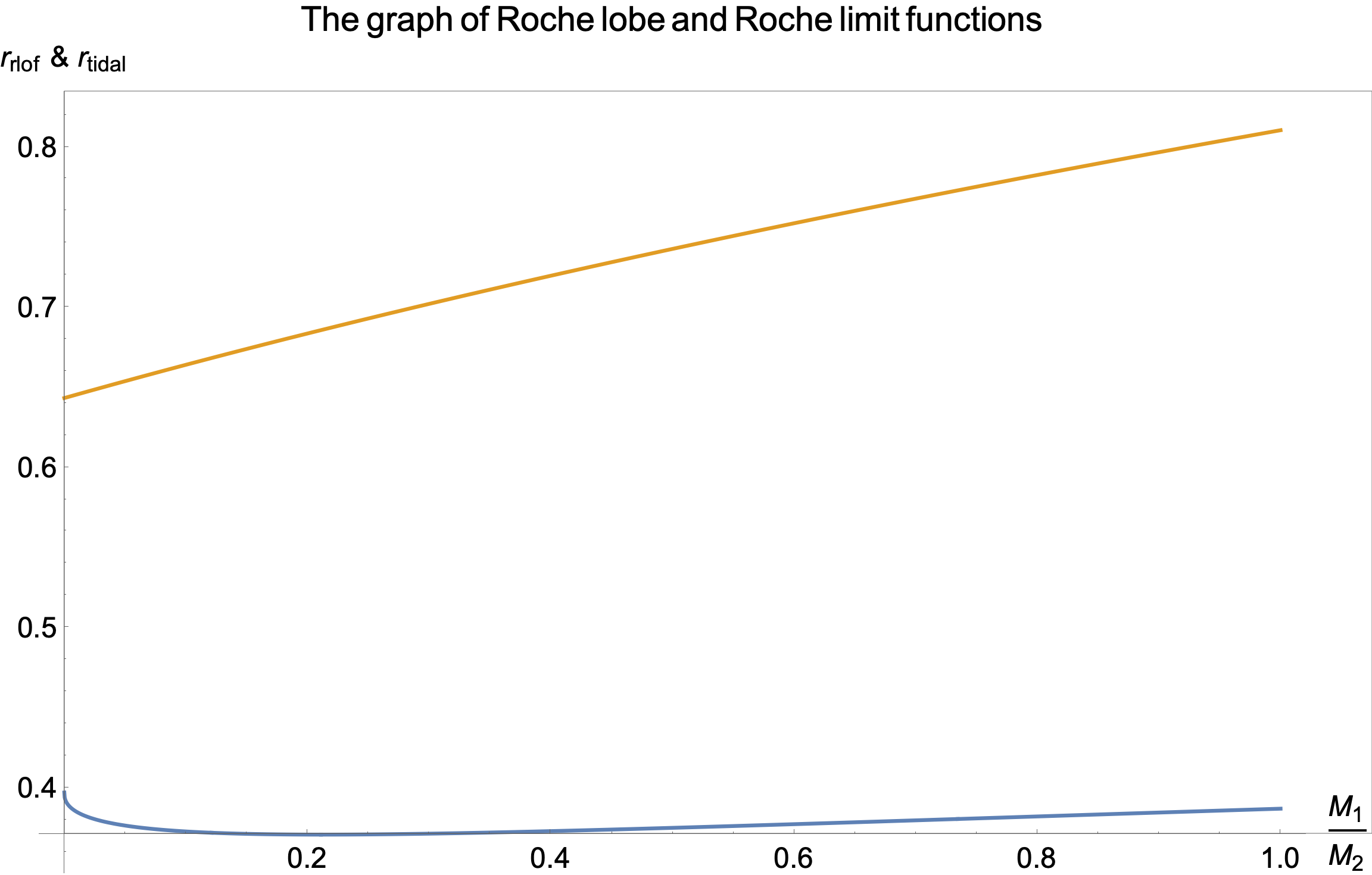}
      \caption{The Roche lobe and Roche limit. This is the diagram of formula \ref{eq:roche} (blue) and \ref{eq:tidal} (orange). The independent variable is the mass ratio $\frac{M_1}{M_2}$, where $M_1$ is fixed to 1~$M_\sun$. The dependent variable are $r_{\text{rlof}}$ (blue) and $r_{\text{tidal}}$ (orange). In the range of mass ratio from 0 to 1, $r_{\text{rlof}}$ never excess Roche limit ($r_{\text{tidal}}$) and X-24 is not destroyed by the tidal force.}
      \label{fig6}
      \end{center}
   \end{figure}  
We also tried to use the Roche limit ($r_{\rm tidal}$) to put constraints on the mass of the compact object. As mentioned, the system is likely a compact binary with RLOF. If the compact object is very massive so that $r_{\rm rlof}>r_{\rm tidal}$, the tidal force around the compact object will tear the low-mass companion apart (i.e., a tidal disruption event) and we should not see a stable system like X-24 as \textit{Chandra} observed. The Roche limit can be written as \cite{1893Natur..48...54D}.%\citep{1893Natur..48...54D}.

\begin{equation}
r_{\rm tidal}=a(\frac{M_1}{2M_2})^{\frac{1}{3}}.
\label{eq:tidal}
\end{equation}
Assuming $P_{\rm orb}=7500$s and $M_1=1{M}_\sun$, we find that $r_{\rm tidal}$ is always larger than $r_{\rm rlof}$ for $M_2<100M_\sun$. In other words, the Roche limit does not constrain the mass of the compact object for X-24. From the point of view of the actual star, such as if the stellar companion is a giant, under the condition of the Roche limit, the star will be affected by the tidal force to disintegrate before growing into a giant star. Therefore, the possibility of the compact object engulfed in the atmosphere of the giant star is not high.

%--------------------------------------------------------------------
%\begin{table}[h]         % title of Table      % is used to refer this table in the text
%\caption{The table of RLOF and RL with the different mass of BH. \textbf{(Ray: we can merge this table to Figure 6)}\\}
%\label{table:1}
%\centering 
%\tabcolsep=6pt
%\renewcommand\arraystretch{1.3}                    % used for centering table
%\begin{tabular}{|c | c | c | c|}        % centered columns (4 columns)
%\hline                 % inserts double horizontal lines
%The BH's mass & Distance & RLOF & The maximal radius \\ 
% (\(\textup{M}_\odot\)) & ( \(\textup{R}_\odot\) )&  (\(\textup{R}_\odot\)) &   (\(\textup{R}_\odot\)) \\      % table heading 
%\hline                        % inserts single horizontal line
%   1 & 1.02 & 0.3870 & 0.81 \\      % inserting body of the table
%\hline 
%   10 & 1.80 & 0.3727    & 0.66 \\
%\hline 
%   100 & 3.77 & 0.3851   & 0.65 \\
%\hline                                   %inserts single line
%\end{tabular}
%\tablecomments{The maximal radius of fourth column is estimated from the RL.}
%\end{table}
%----------------------------------------------------------------- 

Although both the RLOF and Roche limit considerations do not give constraints on the mass of the compact object, an IMBH scenario is unlikely for X-24. If X-24 is an IMBH system, the formation channel of the binary will probably be tidal capture \citep{2006MNRAS.372..467B}. However, the orbital period of X-24 is short, and hence, the stellar separation is small. The chance that the companion star got captured by the IMBH within this tiny region is therefore very low. On the contrary, if the compact star's mass is similar to that of its companion, the two stars were likely born together and then evolved to form the binary system observed. If the speculation is correct, X-24 will be either a pulsar or a stellar-mass BH. Pulsars in compact binaries are well known as black widow and redback millisecond pulsar binaries \citep{2019Galax...7...93H}, but their X-ray luminosities are much lower than $10^{39}$\lum\ \citep{2018ApJ...864...23L}. Therefore, a stellar-mass BH scenario with super-Eddington accretion is favoured for X-24.

\section{Conclusions} \label{sec:cite}

We report a study of the X-ray population in NGC~1559 using the \textit{Chandra}/ACIS observation taken in 2016. 33 X-ray point sources were detected. Among these, there are 8 ULXs with X-ray luminosities higher than $10^{39}$\lum. Considering the contamination of background, the number of bonafide ULXs is almost 7, which is an unexpectedly large number. We have tried to explain the excess of ULXs using the arguments of low-metallicity or starburst and compared with similar cases of other nearby galaxies. Unfortunately, there is still no satisfactory explanation.

In these ULXs, most are non-thermal systems except for X-6 that has a photon index larger than 2.5, implying a possible thermal origin of the emission. Besides spectral analyses, we examined the X-ray light curves of the sources, and found that the ULX, X-24, showed an X-ray periodicity of 7500s with a detection significance of $\gtrapprox2.7\sigma$ (two tailed). Assuming that it is a LMXB, we attempted to constrain the mass of the compact object with the RLOF and Roche limit cases, but the results cannot rule out the possibility that it is an IMBH system. Nevertheless, By considering the formation channel of the compact binary and comparing the observed X-ray luminosity of X-24 to some well-known cases of with neutron star compact binaries (also known as black widow and redback pulsars), we favor a stellar-mass BH scenario for X-24. In any case, the mechanism of the ULXs with a 2 hour period is not clear due to the lack of observational data. It is worth carrying out further follow-up observations to reveal its true physical origin.

\begin{acknowledgments}
This research has made use of data obtained from the \textit{Chandra} Data Archive and the \textit{Chandra} Source Catalog, and software provided by the \textit{Chandra} X-ray Center (CXC) in the application packages \texttt{CIAO} and \texttt{Sherpa}.
C.H.M and K.L.L. are supported by the Ministry of Science and Technology of the Republic of China (Taiwan) through grant 110-2636-M-006-013. K.L.L is a Yushan (Young) Scholar of the Ministry of Education of the Republic of China (Taiwan).
Y.H.C. acknowledges the support of NSTC grant 110-2112-M-001-020 from the National Science and Technology Council of Taiwan. A.K.H.K. is supported by the National Science and Technology Council of Taiwan through grant 111-2112-M-007-020.
\end{acknowledgments}




%%\appendix

%%\section{Appendix information}


\bibliography{sample631}{}
\bibliographystyle{aasjournal}



\end{document}